\documentclass[usletter,journal]{IEEEtran}
\usepackage[utf8]{inputenc}
\usepackage[T1]{fontenc}
\usepackage{array}
\usepackage{colortbl}
\usepackage[caption=false]{subfig}
\usepackage{tabularx}
\usepackage{tikz}
\usepackage{xcolor}
\usepackage{xspace}
\usepackage{amssymb}
\usepackage{amsmath}
\usepackage{courier}
\usepackage{manfnt}
\usepackage{stmaryrd}
\usepackage{url}
\usepackage{cite}

\font\realtt=cmtt10
\newcommand{\code}[1]{\ensuremath{\text{\realtt #1}}}
\newcommand{\codemu}[1]{\code{MU[$#1$]}}
\newcommand{\codecnt}[1]{\code{CNT[$#1$]}}

\newcommand{\bbbn}{\ensuremath{\text{I\!N}}}
\newcommand{\bbbb}{\ensuremath{\text{I\!B}}}
\newcommand{\bfunc}{\ensuremath{\mathcal{B}}}
\newcommand{\defeq}{\stackrel{\text{\normalfont def}}{=}}

\newcommand{\onset}{\ensuremath{\operatorname{on}}}
\newcommand{\offset}{\ensuremath{\operatorname{off}}}
\newcommand{\inc}{\ensuremath{\operatorname{inc}}}
\newcommand{\cube}{\ensuremath{\operatorname{cube}}}
\newcommand{\dc}{\ensuremath{\operatorname{dc}}}
\newcommand{\constant}{\ensuremath{\kappa}}  

\tikzset{%
  terminal vertex/.style={draw,rectangle,inner sep=0pt,minimum width=10pt,minimum height=10pt},
}
\def\low{\mathop{\rm low}}
\def\high{\mathop{\rm high}}

\def\bddFalse{\tikz[baseline=(x.base)] \node[terminal vertex] (x) {$\bot$};}
\def\bddTrue{\tikz[baseline=(x.base)] \node[terminal vertex] (x) {$\top$};}

\newtheorem{proposition}{Proposition}
\newtheorem{lemma}{Lemma}
\newtheorem{theorem}{Theorem}

\newtheorem{example}{Example}
\newtheorem{remark}{Remark}

\newcommand{\coNP}{\ensuremath{\mathsf{coNP}}\xspace}

\def\xskip{\hskip 7pt plus 3pt minus 4pt}

\newdimen\algindent
\newif\ifitempar \itempartrue 
\def\algindentset#1{\setbox0\hbox{{\bf #1.\kern.25em}}\algindent=\wd0\relax}
\def\algbegin #1 #2{\algindentset{#21}\alg #1 #2} 
\def\aalgbegin #1 #2{\algindentset{#211}\alg #1 #2} 
\def\alg#1(#2). {\medbreak 
  \noindent{\bf#1}({\it#2\/}).\xskip\ignorespaces}
\def\algstep#1.{\ifitempar\smallskip\noindent\else\itempartrue
  \hskip-\parindent\fi
  \hbox to\algindent{\bf\hfil #1.\kern.25em}%
  \hangindent=\algindent\hangafter=1\ignorespaces}
\def\slug{\hbox{\kern1.5pt\vrule width2.5pt height6pt depth1.5pt\kern1.5pt}}

\newcolumntype{t}{>{\ttfamily}c}

\def\hang{\hangindent19pt}
\def\d@anger{\medbreak\begingroup\clubpenalty=10000
 \def\par{\endgraf\endgroup\medbreak} \noindent\hang\hangafter=-2
 \hbox to0pt{\hskip-\hangindent\dbend\hfill}\small}
\outer\def\danger{\d@anger}

\usetikzlibrary{calc,decorations.markings,shapes.geometric,patterns}

\newcolumntype{C}{>{$}c<{$}}
\newcolumntype{H}{>{\highlight}C}

\tikzset{%
  font=\footnotesize,>=stealth,
  bdd vertex/.style={draw,circle,inner sep=0pt,minimum size=12pt},
  qmdd vertex/.style={draw,ellipse,inner sep=0pt,minimum height=10pt},
  terminal vertex/.style={draw,rectangle,inner sep=0pt,minimum width=10pt,minimum height=10pt},
  qmdd zero/.style={inner sep=0pt,below,font=\scriptsize}
}

\def\tikzAndEmbedding{%
  \def\highlight{}
  \def\p{\phantom{0}}
  \def\pl{\multicolumn{5}{c}{\p} \\}
  \begin{tabular}{HH|HCC}
    \hline
    x_1 & x_2 & y & \gamma_1 & \gamma_2\\
    \noalign{\gdef\highlight{\bf}}
    \hline
    0 & 0 & 0 & 0 & 0 \\
    0 & 1 & 0 & 0 & 1 \\
    1 & 0 & 0 & 1 & 0 \\
    1 & 1 & 1 & 0 & 0 \\
    \hline
    \pl
    \pl
    \pl
    \pl
  \end{tabular}
}

\def\tikzAndEmbeddingConstants{%
  \def\highlight{}
  \begin{tabular}{CHH|HCC}
    \hline
    \constant & x_1 & x_2 & y & \gamma_1 & \gamma_2\\
    \noalign{\gdef\highlight{\bf}}
    \hline
    0 & 0 & 0 & 0 & 0 & 0 \\
    0 & 0 & 1 & 0 & 0 & 1 \\
    0 & 1 & 0 & 0 & 1 & 0 \\
    0 & 1 & 1 & 1 & 0 & 0 \\
    \noalign{\gdef\highlight{}}
    1 & 0 & 0 & 0 & 1 & 1 \\
    1 & 0 & 1 & 1 & 0 & 1 \\
    1 & 1 & 0 & 1 & 1 & 1 \\
    1 & 1 & 1 & 1 & 1 & 1 \\
    \hline
  \end{tabular}
}

\newcommand\tikzUnfoldingSubsets{%
\begin{tikzpicture}
  \begin{scope}
    \clip (-.75,0) ellipse [x radius=1.5cm,y radius=.6cm];
    \fill[pattern color=gray!50!white,pattern=north east lines] (.75,0) ellipse [x radius=1.5cm,y radius=.6cm];
  \end{scope}

  \begin{scope}[every node/.style={draw,ellipse,minimum width=3cm,minimum height=1.2cm}]
    \node (p) at (-.75,0) {};
    \node (pp) at (.75,0) {};
  \end{scope}
  \node at (0,0) {$c\land c'$};
  \node at (-1.3,0) {$c\land \bar c'$};
  \node at (1.3,0) {$c'\land \bar c$};

  \node[below,xshift=-10pt] at (p.south) {$e=(c,o)$};
  \node[below,xshift=10pt] at (pp.south) {$e'=(c',o')$};

  \begin{scope}[every node/.style={above,inner sep=.5pt}]
    \draw[->] (-1.8,.3) to ++(up:15pt) node[xshift=-15pt] {$P_f\leftarrow P_f\cup\{(c\land\bar c',o)\}$};
    \draw[->] (1.8,.3) to ++(up:15pt) node[xshift=15pt] {$P'_f\leftarrow P'_f\cup\{(c'\land\bar c,o')\}$};
    \draw[->] (0,.3) to ++(up:35pt) node {$P'_f\leftarrow P'_f\cup\{(c\land c',o\cup o')\}$};
  \end{scope}
\end{tikzpicture}
}

\def\tikzHeuristicEmbeddingScheme{%
  \def\!{\cellcolor{gray!10}}
  \begin{tabular}{>{$}c<{$}>{$}c<{$}|>{$}c<{$}>{$}c<{$}} \hline
    c_1\dots c_m & x_1 \dots x_n & y_1\dots y_m & \gamma_1 \dots \gamma_n \\\hline
    0\dots00 & \!0 \dots 0 & \!f_1(\vec x)\dots f_m(\vec x) & 0 \dots 0 \\
    \vdots & \!\vdots & \!\vdots & \vdots \\
    0\dots00 & \!1 \dots 1 & \!f_1(\vec x)\dots f_m(\vec x) & 1 \dots 1 \\
    0\dots01 & 0 \dots 0 & f_1(\vec x)\dots\overline{f_m(\vec x)} & 0 \dots 0 \\
    \vdots & \vdots & \vdots & \vdots \\
    0\dots01 & 1 \dots 1 & f_1(\vec x)\dots\overline{f_m(\vec x)} & 1 \dots 1 \\
    \vdots & \vdots & \vdots & \vdots \\
    1\dots11 & 0 \dots 0 & \overline{f_1(\vec x)}\dots \overline{f_m(\vec x)} & 0 \dots 0 \\
    \vdots & \vdots & \vdots & \vdots \\
    1\dots11 & 1 \dots 1 & \overline{f_1(\vec x)}\dots \overline{f_m(\vec x)} & 1 \dots 1 \\\hline
  \end{tabular}
}

\def\tikzBDDandQMDD{%
\begin{tikzpicture}
  \begin{scope}[every node/.style={bdd vertex}]
    \node (x) {$x$};
    \node (y left) at ($(x)+(-125:.8cm)$) {$y$};
    \node (y right) at ($(x)+(-55:.8cm)$) {$y$};
  \end{scope}

  \node (f0) at ($(y left)+(-110:.8cm)$) {$f$};
  \node (f3) at ($(y right)+(-70:.8cm)$) {$f'''$};
  \node (f1) at ($(y left)+(-70:.8cm)$) {$f'$};
  \node (f2) at ($(y right)+(-110:.8cm)$) {$f''$};

  \draw (x) -- (y right) -- (f3) (y left) -- (f1);
  \draw[dashed] (x) -- (y left) -- (f0) (y right) -- (f2);

  \begin{scope}[xshift=3cm]
    \begin{scope}[every node/.style={qmdd vertex}]
      \node (xy) {$x\mapsto y$};
      \node[opacity=0] (y left) at ($(xy)+(-125:.8cm)$) {$y$};
      \node[opacity=0] (y right) at ($(xy)+(-55:.8cm)$) {$y$};
    \end{scope}

    \node (f0) at ($(y left)+(-110:.8cm)$) {$f$};
    \node (f3) at ($(y right)+(-70:.8cm)$) {$f'''$};
    \node (f2) at ($(y left)+(-70:.8cm)$) {$f''$};
    \node (f1) at ($(y right)+(-110:.8cm)$) {$f'$};

    \draw (xy) -- +(225:.8cm) -- (f0);
    \draw (xy) -- +(255:.8cm) -- (f2);
    \draw (xy) -- +(285:.8cm) -- (f1);
    \draw (xy) -- +(315:.8cm) -- (f3);
  \end{scope}

  \draw[shorten <=20pt,shorten >=20pt,->] (x) -- (xy);
\end{tikzpicture}
}

\def\tikzVariableOrdering{%
  \begin{tikzpicture}
    \def\r{1.05cm}
    \draw (90:\r) -- (210:\r) -- (330:\r) -- cycle;
    \node {$\kappa_i$'s/$x_j$'s};

    \coordinate (l) at (210:\r);
    \coordinate (r) at (330:\r);
    \coordinate (a) at ([yshift=-.4cm] l);
    \coordinate (b) at ([yshift=-.4cm] r);

    \node[bdd vertex] at ([xshift=-6pt] a) (ya) {$y_1$};
    \node[bdd vertex] at ([xshift=-2pt] $(a)!1/3!(b)$) (yb) {$y_1$};
    \node[bdd vertex] at ([xshift=2pt] $(a)!2/3!(b)$) (yc) {$y_1$};
    \node[bdd vertex] at ([xshift=6pt] b) (yd) {$y_1$};

    \node at ([xshift=1pt] $(yc)!.5!(yd)$) {\dots};

    \draw ($(l)!1/5!(r)$) -- (ya);
    \draw ($(l)!2/5!(r)$) -- (yb);
    \draw ($(l)!3/5!(r)$) -- (yc);
    \draw ($(l)!4/5!(r)$) -- (yd);

    \coordinate (c) at ($(210:\r)!2!(a)$);
    \coordinate (d) at ($(330:\r)!2!(b)$);

    \draw (ya) -- ($(c)!.15!(d)$);
    \draw (ya) -- ($(c)!.25!(d)$);
    \draw (yb) -- ($(c)!.35!(d)$);
    \draw (yb) -- ($(c)!.45!(d)$);
    \draw (yc) -- ($(c)!.55!(d)$);
    \draw (yc) -- ($(c)!.65!(d)$);
    \draw (yd) -- ($(c)!.75!(d)$);
    \draw (yd) -- ($(c)!.85!(d)$);

    \draw (c) -- (d) -- ++(-60:.5cm) -- ++(-120:.5cm) coordinate (e) -- (e -| a) coordinate (f) -- ++(120:.5cm) -- ++(60:.5cm);
    \node at ($(c)!.5!(e)$) {$y_i$'s/$\gamma_j$'s};

    \coordinate (g) at ($(f)!1/3!(e)$);
    \coordinate (h) at ($(f)!2/3!(e)$);
    \draw (g) -- ++(down:.5cm) node[below,terminal vertex] (zero) {$\bot$} -- (h);
    \draw (h) -- ++(down:.5cm) node[below,terminal vertex] (one) {$\top$} -- (g);

    \draw[dashed] ([xshift=-3pt,yshift=3pt] ya.north west) rectangle ([xshift=3pt,yshift=-3pt] yd.south east);
    \node[right] at ([yshift=2pt] yd.east) {\enspace$\leftarrow 2^{n+m}$ vertices};
  \end{tikzpicture}
}

\newcommand{\tikzCharBDDExampleWithZeros}{%
\begin{tikzpicture}[scale=.25]
\begin{scope}[thin,every node/.style={draw,ellipse,inner sep=1.5pt},low edge/.style={dashed},decoration={markings,mark=at position .5 with {\fill (0,0) circle (1.5pt);}}]
  \node (x2-5) at (911bp,306bp) [] {$x_2$};
  \node (y3-3) at (864bp,450bp) [] {$y_3$};
  \node (x3-3) at (909bp,234bp) [] {$x_3$};
  \node (y1-1) at (520bp,594bp) [] {$y_1$};
  \node[fill=gray] (x1-5) at (1002bp,378bp) [] {$x_1$};
  \node (x3-1) at (250bp,234bp) [] {$x_3$};
  \node (x2-3) at (690bp,306bp) [] {$x_2$};
  \node[fill=gray] (x1-3) at (691bp,378bp) [] {$x_1$};
  \node (x2-1) at (279bp,306bp) [] {$x_2$};
  \node[fill=gray] (x1-1) at (279bp,378bp) [] {$x_1$};
  \node (x4-4) at (523bp,162bp) [] {$x_4$};
  \node (x2-4) at (819bp,306bp) [] {$x_2$};
  \node (y2-1) at (354bp,522bp) [] {$y_2$};
  \node (x4-1) at (250bp,162bp) [] {$x_4$};
  \node[fill=gray] (x1-2) at (400bp,378bp) [] {$x_1$};
  \node (x4-3) at (432bp,162bp) [] {$x_4$};
  \node (x2-2) at (400bp,306bp) [] {$x_2$};
  \node (x3-2) at (432bp,234bp) [] {$x_3$};
  \node[draw=none] (f) at (520bp,696bp) [] {$f$};
  \node (x2-6) at (1002bp,306bp) [] {$x_2$};
  \node (x4-2) at (341bp,162bp) [] {$x_4$};
  \node (y3-1) at (354bp,450bp) [] {$y_3$};
  \node (y3-2) at (658bp,450bp) [] {$y_3$};
  \node[fill=gray] (x1-4) at (864bp,378bp) [] {$x_1$};
  \node[draw,rectangle] (top) at (607bp,18bp) [] {$\top$};
  \node (x5-1) at (446bp,90bp) [] {$x_5$};
  \node (y2-2) at (658bp,522bp) [] {$y_2$};
  \draw [] (x5-1) ..controls (499.31bp,62.613bp) and (549.06bp,41.058bp)  .. (top);
  \draw [low edge,postaction={decorate}] (x5-1) ..controls (504.52bp,66.654bp) and (550.53bp,46.74bp)  .. (top);
  \draw [low edge] (x1-3) ..controls (690.6bp,348.85bp) and (690.4bp,334.92bp)  .. (x2-3);
  \draw [] (x1-1) ..controls (279bp,348.85bp) and (279bp,334.92bp)  .. (x2-1);
  \draw [] (x4-2) ..controls (348.15bp,123.52bp) and (366.12bp,90.27bp)  .. (392bp,72bp) .. controls (449.62bp,31.321bp) and (535.93bp,21.84bp)  .. (top);
  \draw [low edge,postaction={decorate}] (x4-2) ..controls (365.52bp,124.7bp) and (383.66bp,90.593bp)  .. (410bp,72bp) .. controls (462.64bp,34.835bp) and (539.23bp,23.711bp)  .. (top);
  \draw [low edge,postaction={decorate}] (y2-2) ..controls (726.68bp,497.66bp) and (795.41bp,474.31bp)  .. (y3-3);
  \draw [] (x1-4) ..controls (846.16bp,349.25bp) and (836.75bp,334.6bp)  .. (x2-4);
  \draw [] (x2-4) ..controls (737.55bp,265.93bp) and (604.53bp,202.11bp)  .. (x4-4);
  \draw [] (x2-1) ..controls (267.42bp,277.05bp) and (261.5bp,262.76bp)  .. (x3-1);
  \draw [] (x3-1) ..controls (250bp,204.85bp) and (250bp,190.92bp)  .. (x4-1);
  \draw [low edge,postaction={decorate}] (x2-2) ..controls (381.65bp,260.82bp) and (359.34bp,207.14bp)  .. (x4-2);
  \draw [] (y2-2) ..controls (658bp,492.85bp) and (658bp,478.92bp)  .. (y3-2);
  \draw [] (x2-3) ..controls (675.21bp,248.77bp) and (662.16bp,149.1bp)  .. (631bp,72bp) .. controls (625.88bp,59.338bp) and (617.81bp,46.301bp)  .. (top);
  \draw [low edge,postaction={decorate}] (x2-3) ..controls (693.21bp,248.77bp) and (680.16bp,149.1bp)  .. (649bp,72bp) .. controls (643.88bp,59.338bp) and (635.81bp,46.301bp)  .. (top);
  \draw [low edge,postaction={decorate}] (x1-5) ..controls (1002bp,348.85bp) and (1002bp,334.92bp)  .. (x2-6);
  \draw [] (x4-1) ..controls (286.31bp,125.45bp) and (325.72bp,90.738bp)  .. (366bp,72bp) .. controls (438.63bp,38.207bp) and (533.75bp,25.504bp)  .. (top);
  \draw [low edge,postaction={decorate}] (x1-4) ..controls (882.39bp,349.62bp) and (892.29bp,334.87bp)  .. (x2-5);
  \draw [low edge] (x3-1) ..controls (221.93bp,208.17bp) and (210.41bp,194.73bp)  .. (205bp,180bp) .. controls (199.48bp,164.98bp) and (197.61bp,158.19bp)  .. (205bp,144bp) .. controls (230.81bp,94.468bp) and (256.4bp,93.377bp)  .. (308bp,72bp) .. controls (402.73bp,32.755bp) and (526.11bp,22.55bp)  .. (top);
  \draw [low edge] (x1-1) ..controls (216.78bp,343.83bp) and (148bp,297.22bp)  .. (148bp,235bp) .. controls (148bp,235bp) and (148bp,235bp)  .. (148bp,161bp) .. controls (148bp,107.52bp) and (181.7bp,96.976bp)  .. (229bp,72bp) .. controls (290.31bp,39.624bp) and (503.94bp,24.765bp)  .. (top);
  \draw [low edge,postaction={decorate}] (x2-6) ..controls (991.88bp,244.29bp) and (965.6bp,128.14bp)  .. (893bp,72bp) .. controls (852.65bp,40.8bp) and (697.91bp,25.839bp)  .. (top);
  \draw [] (y3-3) ..controls (864bp,420.85bp) and (864bp,406.92bp)  .. (x1-4);
  \draw [] (x1-5) ..controls (1028.2bp,351.94bp) and (1040bp,338.3bp)  .. (1047bp,324bp) .. controls (1064.9bp,287.71bp) and (1066bp,275.45bp)  .. (1066bp,235bp) .. controls (1066bp,235bp) and (1066bp,235bp)  .. (1066bp,161bp) .. controls (1066bp,70.079bp) and (733.15bp,31.035bp)  .. (top);
  \draw [low edge] (x2-1) ..controls (298.29bp,260.82bp) and (321.73bp,207.14bp)  .. (x4-2);
  \draw [low edge] (y3-3) ..controls (913.64bp,423.82bp) and (952.02bp,404.35bp)  .. (x1-5);
  \draw [] (x3-2) ..controls (432bp,204.85bp) and (432bp,190.92bp)  .. (x4-3);
  \draw [low edge] (x4-1) ..controls (284.07bp,148.34bp) and (290.23bp,146.08bp)  .. (296bp,144bp) .. controls (338.08bp,128.79bp) and (386.97bp,111.61bp)  .. (x5-1);
  \draw [] (x1-2) ..controls (400bp,348.85bp) and (400bp,334.92bp)  .. (x2-2);
  \draw [] (x4-4) ..controls (493.38bp,134.07bp) and (475.58bp,117.89bp)  .. (x5-1);
  \draw [] (y3-2) ..controls (642.35bp,405.94bp) and (626bp,353.25bp)  .. (626bp,307bp) .. controls (626bp,307bp) and (626bp,307bp)  .. (626bp,161bp) .. controls (626bp,115.73bp) and (616.4bp,63.058bp)  .. (top);
  \draw [] (x1-3) ..controls (718.46bp,352.06bp) and (729.72bp,338.62bp)  .. (735bp,324bp) .. controls (740.43bp,308.95bp) and (737.33bp,303.83bp)  .. (735bp,288bp) .. controls (720.4bp,188.82bp) and (725.1bp,155.76bp)  .. (670bp,72bp) .. controls (660.65bp,57.79bp) and (646.38bp,45.339bp)  .. (top);
  \draw [] (x2-6) ..controls (966.77bp,278.48bp) and (944.2bp,261.49bp)  .. (x3-3);
  \draw [low edge,postaction={decorate}] (x2-4) ..controls (824.64bp,245.45bp) and (828.11bp,134.76bp)  .. (772bp,72bp) .. controls (736.51bp,32.299bp) and (671.22bp,22.161bp)  .. (top);
  \draw [low edge] (y1-1) ..controls (569.64bp,567.82bp) and (608.02bp,548.35bp)  .. (y2-2);
  \draw [low edge] (y3-1) ..controls (325.15bp,422.07bp) and (307.81bp,405.89bp)  .. (x1-1);
  \draw [low edge] (y3-2) ..controls (671.18bp,421.05bp) and (677.91bp,406.76bp)  .. (x1-3);
  \draw [] (x2-2) ..controls (412.78bp,277.05bp) and (419.31bp,262.76bp)  .. (x3-2);
  \draw [] (y2-1) ..controls (251.57bp,506.94bp) and (110bp,474.57bp)  .. (110bp,379bp) .. controls (110bp,379bp) and (110bp,379bp)  .. (110bp,161bp) .. controls (110bp,114.27bp) and (126.51bp,96.998bp)  .. (166bp,72bp) .. controls (235.39bp,28.071bp) and (494.6bp,20.552bp)  .. (top);
  \draw [] (x4-3) ..controls (430.61bp,132.68bp) and (433.47bp,118.5bp)  .. (x5-1);
  \draw [low edge,postaction={decorate}] (x4-3) ..controls (444.51bp,133.58bp) and (447.36bp,119.45bp)  .. (x5-1);
  \draw [low edge] (y2-1) ..controls (354bp,492.85bp) and (354bp,478.92bp)  .. (y3-1);
  \draw [] (x3-3) ..controls (898.36bp,182.94bp) and (890.32bp,110.16bp)  .. (848bp,72bp) .. controls (816.2bp,43.332bp) and (690.25bp,27.766bp)  .. (top);
  \draw [low edge,postaction={decorate}] (x3-3) ..controls (916.36bp,182.94bp) and (908.32bp,110.16bp)  .. (866bp,72bp) .. controls (832.39bp,41.698bp) and (693.59bp,26.034bp)  .. (top);
  \draw [low edge] (x3-2) ..controls (466.37bp,206.56bp) and (488.26bp,189.72bp)  .. (x4-4);
  \draw [low edge,postaction={decorate}] (f) ..controls (520bp,636.85bp) and (520bp,622.92bp)  .. (y1-1);
  \draw [] (y3-1) ..controls (372.23bp,421.25bp) and (381.86bp,406.6bp)  .. (x1-2);
  \draw [low edge,postaction={decorate}] (x4-4) ..controls (542.3bp,126.55bp) and (559.02bp,97.071bp)  .. (574bp,72bp) .. controls (581.23bp,59.904bp) and (589.62bp,46.426bp)  .. (top);
  \draw [low edge] (x1-2) ..controls (485.53bp,352.86bp) and (588bp,312.31bp)  .. (588bp,235bp) .. controls (588bp,235bp) and (588bp,235bp)  .. (588bp,161bp) .. controls (588bp,115.73bp) and (597.6bp,63.058bp)  .. (top);
  \draw [low edge] (x2-5) ..controls (884.36bp,279.71bp) and (872.12bp,266.06bp)  .. (864bp,252bp) .. controls (821.4bp,178.21bp) and (861.87bp,128.4bp)  .. (798bp,72bp) .. controls (751.18bp,30.659bp) and (674.74bp,21.297bp)  .. (top);
  \draw [] (y1-1) ..controls (462bp,568.54bp) and (411.59bp,547.29bp)  .. (y2-1);
  \draw [] (x2-5) ..controls (910.2bp,276.85bp) and (909.8bp,262.92bp)  .. (x3-3);

\end{scope}
\node[above left,inner sep=0pt,font=\scriptsize] at (x1-1.north west) {$100$};
\node[above right,inner sep=0pt,font=\scriptsize] at (x1-2.north east) {$101$};
\node[above right,inner sep=0pt,font=\scriptsize] at (x1-3.north east) {$010$};
\node[above left,inner sep=0pt,font=\scriptsize] at (x1-4.north west) {$001$};
\node[above right,inner sep=0pt,font=\scriptsize] at (x1-5.north east) {$000$};
\end{tikzpicture}
}

\title{Embedding of Large Boolean Functions for Reversible Logic}
\author{Mathias Soeken, \IEEEmembership{Member, IEEE,}
        Robert Wille, \IEEEmembership{Member, IEEE,}
        Oliver Keszocze, \IEEEmembership{Student Member, IEEE,}
        D.~Michael Miller, \IEEEmembership{Member, IEEE,}
        Rolf Drechsler, \IEEEmembership{Senior Member, IEEE}
\thanks{This work has been submitted to the IEEE for possible publication.
  Copyright may be transferred without notice, after which this version may no
  longer be accessible.}
\thanks{M.~Soeken, R.~Wille, O.~Keszocze, and R.~Drechsler are with the
  University of Bremen, Germany, and the German Research Center for Artificial
  Intelligence (DFKI).}
\thanks{D.~M.~Miller is with the University of Victoria, BC, Canada.}
\thanks{M.~Soeken is corresponding author: msoeken@cs.uni-bremen.de}
}

\begin{document}

\maketitle

\begin{abstract}
  Reversible logic represents the basis for many emerging technologies and has
  recently been intensively studied.  However, most of the Boolean functions of
  practical interest are irreversible and must be embedded into a reversible 
  function before they can be
  synthesized.  Thus far, an optimal embedding is guaranteed only for small
  functions, whereas a significant overhead results when large functions
  are considered. In this paper, we study this issue. We prove that determining an
  optimal embedding is \coNP-hard already for restricted cases.  Then, we
  propose heuristic and exact methods for determining both the number of
  additional lines as well as a corresponding embedding.  For the approaches we
  considered sums of products and binary decision diagrams as function
  representations.  Experimental evaluations show the applicability of the
  approaches for large functions.  Consequently, the reversible embedding of
  large functions is enabled as a precursor to subsequent synthesis.
\end{abstract}

\section{Introduction}
\IEEEPARstart{S}{ynthesis} of reversible circuits has been intensively studied
in the recent
past~\cite{DBLP:conf/dac/MillerMD03,journals/tcad/ShendePMH03,DBLP:conf/aspdac/SoekenWHPD12,VR:08,LJ:14}.
Since most Boolean functions of practical interest are irreversible, such
functions are \emph{embedded} into reversible ones prior to synthesis.  Given an
$m$-output irreversible function~$f$ on~$n$ variables, a \emph{reversible}
function~$g$ with $m+k$ outputs is determined such that~$g$ agrees with~$f$ on
the first~$m$ components.  The overhead in terms of the~$k$ additional variables
shall be kept as small as possible.  The embedding is called \emph{optimal} if
$k$~is minimal.

Thus far, only synthesis approaches based on truth tables allow for a
determination of an optimal embedding.  However, determining an efficient
embedding for large irreversible functions, i.e.~functions with up to a hundred
variables, is an open research problem which significantly hindered the
development of scalable synthesis approaches for reversible logic.

In this work, we study this issue from both, theoretical and practical,
perspectives.  First, we derive two lower bounds for determining the minimal
value of~$k$, namely (1)~we show that already when $m=1$ it is $\coNP$-hard to
determine if the minimal~$k$ equals $n-1$ and (2)~we show that, even when $n-m$
is bounded by a constant, it is $\coNP$-hard to decide if the minimal $k$ equals
$n-m$.  Hence, computing the minimal number $k$ of additional variables is not
feasable in polynomial time unless $\mathsf{P}=\mathsf{NP}$.

We then propose algorithms for both heuristic and optimal embeddings and
evaluate for which cases an efficient application is possible.  We differentiate
between (1)~determining the required number of additional lines and~(2)
determining the concrete embedding.  The key element in both steps is to use
sum-of-product expressions~(SOPs) and binary decision diagrams~(BDDs) which
potentially allow for a more compact function representation compared to truth
tables. As a result, an embedding methodology results which can process large
irreversible functions for the first time.

While so far efficient embedding of irreversible functionality was restricted to
very small functions, the proposed approach enables embedding of functions
containing hundreds of variables.  We confirm this by comprehensive experimental
evaluations.

The contributions described in this paper are as follows:
\begin{itemize}
\item We provide lower bounds for the embedding problem.
\item We present three algorithms for determining the number of additional
  lines of large irreversible functions, one heuristic algorithm (cube-based)
  and two exact ones (cube-based and BDD-based).
\item We propose two algorithms for embedding large irreversible functions,
  i.e.~one exact algorithm (cube-based) and one heuristic algorithm (BDD-based) that
  respects the theoretical upper bound.
\item Finally, we provide open source implementations for all presented
  algorithms.
\end{itemize}

The paper is organized as follows.  Preliminary definitions are given in the
next section.  Section~\ref{sec:synthesis-reversible} provides the background on
the synthesis of reversible function and motivates the problem that is addressed
by this work.  We present known upper bounds and derive new lower bounds for the
problem in Section~\ref{sec:bounds}.  Approaches for approximating and
determining the minimal number of additional lines are described in
Section~\ref{sec:calculating-lines}.  Afterwards, approaches for exact and
heuristic approaches are described in Section~\ref{sec:embedding}.
Section~\ref{sec:exper-eval} presents and discusses the results from the
experimental evaluation before the paper is concluded in
Section~\ref{sec:conclusions}.

\section{Preliminaries}\label{sec:prelim}
In this section we introduce notations.  In Section~\ref{sec:bool-revers-funct}
we introduce (reversible) Boolean functions, in
Section~\ref{sec:binary-decis-diagr} we review BDDs, and in
Section~\ref{sec:sum-prod-repr} we define notations for SOPs.

\subsection{Boolean Functions and Reversible Boolean Functions}
\label{sec:bool-revers-funct}
Let~$\bbbb\defeq\{0,1\}$ denote the \emph{Boolean values} and let
\begin{equation}
\bfunc_{n,m}\defeq\{f\mid f\colon\bbbb^n\to\bbbb^m\}
\end{equation}
be the set of all \emph{Boolean functions} with $n$~inputs and $m$~outputs,
where~$m,n\geq1$.
We write~$\bfunc_n\defeq\bfunc_{n,1}$ for each~$n\geq1$ and assume that each
$f\in\bfunc_n$ is represented by a propositional formula over the variables
$\{x_1,\dots,x_n\}$.
Conversely, any \mbox{$m$-tuple} $t$ of Boolean functions over variables $\{x_1,\ldots,x_n\}$
corresponds to a unique Boolean function $f_t\in\bfunc_{n,m}$.
We assume that each function~$f\in \bfunc_{n,m}$ is represented as a tuple
$f=(f_1,\dots,f_m)$ where~$f_i\in\bfunc_n$ for each~$i\in\{1,\dots,m\}$ and
hence~$f(\vec x)=(f_1(\vec x),\dots,f_m(\vec x))$ for each~$\vec x\in\bbbb^n$.

Given a Boolean function~$f\in\bfunc_{n,m}$ the sets
$\onset(f)\defeq\{\vec x\in\bbbb^n\mid f(\vec x)\neq0^m\}$
and
$\offset(f)\defeq\{\vec x\in\bbbb^n\mid f(\vec x)=0^m\}$
are called \emph{ON-set} and \emph{OFF-set} of~$f$.
It can easily be seen that~$\onset(f)\cup\offset(f)=\bbbb^n$.

Assume~$f=(f_1,\dots,f_m)\in\bfunc_{n,m}$ and~$g=(g_1,\dots,g_{m'})\in
\bfunc_{n,m'}$, where~$m'\geq m$.
We write~$f=g\vert_m$ in case~$f_i(\vec x)=g_i(\vec x)$ for
each~$\vec x\in\bbbb^n$ and each~$i\in\{1,\dots,m\}$ and say that~$f$ is
the~\emph{$m$-projection} of $g$.
We say~$f\in\bfunc_n$ is \emph{valid} if~$f(\vec x)=1$ for
each~$\vec x\in\bbbb^n$.

Given a function~$f=(f_1,\dots,f_m)\in\bfunc_{n,m}$ its \emph{characteristic
function}~$\chi_f\in\bfunc_{n+m}$ is defined as
\begin{equation}
  \label{eq:characteristic}
  \chi_f(\vec x,\vec y)\quad\defeq\quad
\begin{cases} 1 & f(\vec x)=\vec y\\
0 & \text{otherwise}
\end{cases}
\end{equation}
for each~$\vec x\in\bbbb^n$ and each~$\vec y\in\bbbb^m$.
The characteristic function allows one to represent any multiple-output function as
a single-output function.
It can be computed from a multiple-output function by adding
to the variables $\{x_1,\ldots,x_n\}$ the
additional output variables~$\{y_1,\dots,y_m\}$:
\begin{equation}
  \label{eq:characteristic_computation}
  \bigwedge_{i=1}^m(y_i\leftrightarrow f_i(x_1,\dots,x_n))
\end{equation}

Given a Boolean function~$f\in\bfunc_n$ over the variables~$X=\{x_1,\dots,x_n\}$
and a variable~$x_i\in X$, we define the \emph{positive
  co-factor}~$f_{x_i}\in\bfunc_{n-1}$ and the \emph{negative co-factor}~$f_{\bar
  x_i}\in\bfunc_{n-1}$ as $f_{x_i}=f(x_1,\dots,x_{i-1},1,x_{i+1},\dots,
x_n)$ and~$f_{\bar x_i}=f(x_1,\dots,x_{i-1},0,x_{i+1},\dots,x_n)$, respectively.

\begin{figure}[t]
  \centering
  \subfloat[Garbage outputs~\label{fig:and-embedding-garbage}]{%
      \tikzAndEmbedding}
  \hfill
  \subfloat[Constant input~\label{fig:and-embedding-constants}]{%
      \tikzAndEmbeddingConstants}
  \caption{Embedding of the AND function}
  \label{fig:and-embedding}
\end{figure}
A function~$f\in\bfunc_{n,m}$ is called \emph{reversible} if~$f$ is bijective,
otherwise it is called \emph{irreversible}.
Clearly, if~$f$ is reversible, then~$n=m$.
A function~$g\in\bfunc_{n,m+k}$ \emph{embeds}~$f\in\bfunc_{n,m}$, if~$g$ is
injective and~$f\equiv g\vert_{m}$.
The function~$g$ is called an \emph{embedding} and the additional~$k$ outputs
of~$g$ are referred to as \emph{garbage outputs} and are denoted
by~$\vec\gamma$ later.
We are interested in those embeddings of $f$, where $k$ is minimal.
Let
\[\mu(f)\defeq\max\{\#f^{-1}(\{\vec y\})\mid\vec y\in\bbbb^m\}\] 
denote the number of
occurrences of the most frequent output pattern.
It is not hard to see that~$\ell(f)\defeq\lceil\log_2\mu(f)\rceil$ is both an upper and a lower
bound (and thus an optimal bound) for~$k$.
Thus, if~$k=\ell(f)$, then the embedding~$g$ is called~\emph{optimal}.
\begin{example}
  The AND function~$\land\in\bfunc_{2}$ can be embedded into a reversible
  function~$g\in\bfunc_{2,3}$ which is illustrated in
  Fig.~\ref{fig:and-embedding-garbage}.
  The most frequent output pattern is~$0$, hence~$\mu=3$.
  The embedding~$g$ is optimal.
\end{example} 

In order to obtain a reversible function for an embedding~$g$, additional input
variables might need to be added.
Bijectivity can readily be achieved, e.g.~by adding
additional inputs such
that~$f$ evaluates to its original values in case these inputs are assigned the
constant value~$0$ and each output pattern that is not in the image of~$g$ is
arbitrary distributed among the new input patterns.
The additional inputs are referred to as~\emph{constant inputs}.
\begin{example}
  A constant input assignment, denoted~$\constant$, for the embedded AND
  function in Fig.~\ref{fig:and-embedding-garbage} is given in
  Fig.~\ref{fig:and-embedding-constants}.
\end{example}
Different algorithms that perform an optimal embedding of irreversible functions
based on their truth table description have been proposed in the
past~\cite{MWD:2009b}.

\subsection{Binary Decision Diagrams}
\label{sec:binary-decis-diagr}
\emph{Binary Decision Diagrams} (BDD)~\cite{journals/toc/Bryant86} are an established data
structure for representing Boo\-lean functions.
While the general concepts are briefly outlined in this section, the reader is
referred to the literature for a comprehensive
over\-view~\cite{journals/toc/Bryant86,TAOCP4A}.

Let~$X=\{x_1,\dots,x_n\}$ be a set of variables of a Boolean function
$f\in\bfunc_n$.  A BDD representing the function $f$ is a directed acyclic graph
$F$ with non-terminal vertices~$N$ and terminal vertices~$T\subseteq\{\bddFalse,
\bddTrue\}$ where $N\cap T=\emptyset$ and~$T\neq\emptyset$.  Each non-terminal
vertex~$v\in N$ is labeled by a variable from $X$ and has exactly two children,
$\low v $ and~$\high v $.  The directed edges to these children are called
\emph{low-edge} and \emph{high-edge} and are drawn dashed and solid,
respectively.  A non-terminal vertex $v$ labeled $x_i$ represents a function
denoted~$\sigma(v)$ given by the \emph{Shannon decomposition}~\cite{Shannon}
\begin{equation}
\label{eq:shannon}
\sigma(v)=\bar x_i\sigma(\low v)+x_i\sigma(\high v)
\end{equation}
\noindent where $\sigma(\low v)$ and $\sigma(\high v)$ are the functions
represented by the children of $v$ with $\sigma(\bddFalse)\equiv0$ and
$\sigma(\bddTrue)\equiv1$.
The BDD $F$ has a single start vertex~$s$ with $\sigma(s)\equiv f$.

A BDD is \emph{ordered} if the variables of the vertices on every path from the
start vertex to a terminal vertex adhere to a specific ordering.  Not all of the
variables need to appear on a particular path and a variable can appear at most
once on any path.  A BDD is \emph{reduced} if there are no two non-terminal
vertices representing the same function, hence the representation of common
subfunctions is shared.  Complemented edges can additionally reduce the size of
a BDD and are marked using a solid dot.  In the following only reduced, ordered
BDDs are considered and for brevity referred to as~BDDs.

Multiple-output functions can be represented by a single BDD that has more than
one start vertex.
Common subfunctions that can be shared among the functions decrease the overall
size of the BDD.
In fact, many practical Boolean functions can efficiently be represented using
BDDs and efficient manipulations and evaluations are
possible~\cite{journals/toc/Bryant86}.

\subsection{Sum-Of-Product Representation}
\label{sec:sum-prod-repr}
Each Boolean function~$f\in\bfunc_{n}$ can be represented in
\emph{Sum-Of-Product}~(SOP) representation in which~$f$ is of the form
\begin{equation}
  \label{eq:sop}
  f(x_1,\dots,x_n)=\bigvee_{i=1}^kx_1^{p_{i,1}}x_2^{p_{i,2}}\cdots x_n^{p_{i,n}}
\end{equation}
with~\emph{polarities}~$p_{i,j}\in\{0,1,2\}$ for~$j\in\{1,\dots,n\}$ and
\begin{equation}
  \label{eq:polarity}
  x^p=
  \begin{cases}
    \bar x & \text{if $p=0$,} \\
    x & \text{if $p=1$,} \\
    1 & \text{if $p=2$.}
  \end{cases}
\end{equation}
If~$p\neq2$, $x^p$ is called a \emph{literal}, otherwise~$x^p$ is referred to
as~\emph{don't care}.  We call each~$c_i=x_1^{p_{i,1}}x_2^{p_{i,2}}\cdots
x_n^{p_{i,n}}$ a~\emph{cube}\footnote{Often, also the term~\emph{monom} is used
  synonymously.} of \emph{weight}
\begin{equation}
  \label{eq:weight}
  \omega(c_i)=\#\{j\mid p_{i,j}\neq2\}.
\end{equation}
That is, the weight refers to the number of literals in~$c_i$.  Note
that~$\#\onset(c_i)=2^{n-\omega(c_i)}$, i.e.~by removing one literal one doubles
the number of input assignments that satisfy~$c_i$.  The set
\begin{equation}
\label{eq:dont-care}
\dc(c)=\{x_i^{p_i}\mid i\in\{1,\dots,n\},p_i=2\}
\end{equation}
refers to all variables that are don't care and hence not contained as literal
in the cube.  One can also represent~$f$ by its cubes~$\{c_1,\dots,c_k\}$ and we
write
\begin{equation}
  \label{eq:is-constructed}
  f\multimap\{c_1,\dots,c_k\}
\end{equation}
where~$\multimap$ reads ``is constructed of.''

\begin{figure}[t]
  \centering
  \def\tabcolsep{2pt}
  \newcolumntype{C}{>{$}c<{$}}
  \begin{tabular}{CCCCC|CCC}
    x_1 & x_2 & x_3 & x_4 & x_5 & y_1 & y_2 & y_3 \\\hline
    1   & -   & -   & 0   & -   & 1   & 0   & 0   \\
    0   & 0   & -   & -   & -   & 0   & 1   & 0   \\
    1   & 1   & -   & -   & 1   & 0   & 0   & 1   \\
    -   & 1   & 0   & -   & -   & 0   & 0   & 1   \\
    1   & 0   & -   & 1   & -   & 1   & 0   & 1   \\
    1   & 1   & -   & 1   & 0   & 1   & 0   & 1   \\
  \end{tabular}
  \caption{PLA representation}
  \label{fig:pla}
\end{figure}

A multiple output function~$f=(f_1,\dots,f_m)\in\bfunc_{n,m}$ is represented
by~$m$ such sum-of-product forms, i.e.~$f_i\multimap C_i$
for~$i\in\{1,\dots,m\}$ where each~$C_i$ is a set of cubes.  All cubes of~$f$
are then given by
\[
  C=\bigcup_{i=1}^mC_i.
\]

Conversely, we can represent~$f$ also as a function that maps each cube from~$C$
to those output functions that are constructed from this cube.  More formally,
$f$ is represented by a function
\begin{equation}
  \label{eq:pla}
  P_f:C\to\mathcal{P}(\{f_1,\dots,f_m\})
\end{equation}
where~$\mathcal{P}$ denotes the power set and with
\[
  P_f(c)=\{f_i\mid c\in C_i\}.
\]
We refer to $P_f$ as the \emph{PLA representation} of~$f$.  Since the power set
of all output functions is being used in several places, we will
use~$\mathcal{P}(f)\defeq\mathcal{P}(\{f_1,\dots,f_m\})$ in the remainder for a
more compact representation.  We will later see that the PLA representation of a
function turns out to be convenient to formulate algorithms.

\begin{example}
  The PLA representation of the function~$f=(f_1,f_2,f_3)\in\bfunc_{5,3}$ with
  \begin{align*}
    y_1 & = x_1\bar x_4 \lor x_1\bar x_2x_4 \lor x_1x_2x_4\bar x_5, \\
    y_2 & = \bar x_1\bar x_2, \\
    y_3 & = x_1x_2x_5 \lor x_2\bar x_3 \lor
    x_1\bar x_2x_4 \lor x_1x_2x_4\bar x_5. \\
  \end{align*}
  is illustrated by the table in Fig.~\ref{fig:pla}.  As before, we make use of
  the convention~$y_i=f_i(x_1,\dots,x_5)$.

  Each input cube of~$f$ is represented by its polarities with the exception
  that we write $-$ instead of $2$.  For each output function we write~$1$, if
  the corresponding cube is in its constructing set, otherwise~$0$.

  The table shall not be confused with a truth table, in particular the~$0$'s in
  the output columns do not indicate that the functions evaluate to~$0$ for the
  corresponding input cube.  As an example, we have~$f_1(x_1,x_2,\bar
  x_3,x_4,\bar x_5)=1$ due to the sixth cube.  However, this input pattern is
  also contained in the fourth cube which is not in the constructing set
  of~$f_1$.
\end{example}

\section{Synthesis of Reversible Functions}
\label{sec:synthesis-reversible}
Reversible circuits on $r$~lines composed of special reversible gates,
e.g.~Toffoli gates, represent reversible functions of $r$~variables.
In turn, every reversible function can be realized with a reversible circuit.
The problem of finding a reversible circuit for a given function is called
synthesis and, since most of the functions of practical interest are
irreversible, we are considering the following synthesis problem in this paper:
Given an arbitrary Boolean function~$f\in\bfunc_{n,m}$,
a reversible circuit should be determined that
represents~$g\in\bfunc_{r,r}$ such that $g$ embeds~$f$.

The synthesis of reversible functions has been an intensively studied research
area in the last decade.
Initially, algorithms based on truth table representations
have been proposed~\cite{DBLP:conf/dac/MillerMD03,
DBLP:journals/tcad/GrosseWDD09}.
Due to their non-scalable representation, none of these is capable of efficiently
synthesizing functions with more than 20~variables.
This fact is reflected in the first two rows in
Table~\ref{tab:synthesis-overview}.
The \emph{R} in the respective rows denotes that these approaches are only
directly applicable for reversible functions.
However, irreversible functions are handled by embedding them first as described
in the previous section.
Moreover, an optimal embedding and therefore a minimal overhead in terms of
additional variables can be guaranteed since the value of $\ell(f)$ can readily
be determined from the truth table, in other words~$r=m+\ell(f)$.
\begin{table}[t]
  \centering
  \def\tabcolsep{1.7pt}
  \caption{Synthesis of reversible functions}
  \label{tab:synthesis-overview}
  \begin{tabularx}{\linewidth}{l>{\em}Xrcc}
    \hline
    \textbf{Representation} & \emph{\textbf{S}} &
    \multicolumn{1}{c}{\textbf{Scalability}} & \textbf{E}&
    \multicolumn{1}{c}{\textbf{Overhead}} \\
    \hline
    Truth table (exact) & R & $\le 7$ variables & y &minimal \\
    Truth table (heuristic) & R & $\le 20$ variables& y & minimal \\
    Symbolic & I & $\approx 100$ variables & N/A&large \\
    Symbolic (QMDDs) & R & $\approx 100$ variables& n & minimal \\
    \hline
  \end{tabularx}

  \smallskip
  S: Supports reversible (R) or also irreversible (I) functions\\
  E: Embedding approaches are available (y) or not (n)
\end{table}

In order to synthesize larger functions, researchers have been investigating the
use of symbolic representations.
An approach based on BDDs~\cite{DBLP:conf/dac/WilleD09} is one of the
first solutions able to synthesize functions with more than one hundred
variables (as summarized in the third row of
Table~\ref{tab:synthesis-overview}).
In particular, by using BDDs it is also possible to directly start with the
irreversible function representation as the embedded takes place implicitly
during synthesis (denoted by~\emph{I} in the second column of
Table~\ref{tab:synthesis-overview}).
However, the newly achieved scalability is traded off against a large number of
additional variables which is much larger than the upper
bound~\cite{conf/date/WilleKD11}, in other words~$r\gg m+n$.

The large number of additional variables results from the fact that this
algorithm performs synthesis in a hierarchical fashion rather than considering
the function as a whole.
Determining an optimal embedding similar to the truth table-based approaches is
not applicable.
Although optimization approaches exist that reduce additional variables in a
post-synthesis step~\cite{DBLP:conf/dac/WilleSD10}, a satisfying result can
rarely be achieved as evaluated in~\cite{conf/date/WilleKD11}.

Following these considerations, an alternative has been proposed
in~\cite{DBLP:conf/aspdac/SoekenWHPD12} that exploits another symbolic
representation and relies on reversible functions.  For this purpose,
\emph{Quantum Multiple-values Decision Diagrams} QMDDs
\cite{DBLP:conf/ismvl/MillerT06} are utilized which are data-structures
particularly suited for the representation of reversible functions.  For the
first time, this enabled the synthesis of reversible functions with up to 100
variables and without adding any additional variables.  However, this algorithm
is of not much help, since so far it was unknown how to embed the given
irreversible function into a QMDD representing the reversible embedding (as
summarized in the forth row of Table~\ref{tab:synthesis-overview}).

In summary, synthesis of reversible circuits for irreversible functions always
requires an efficient embedding.
For small functions, this is no problem and in fact, optimal embeddings can
readily be obtained.
In contrast, if larger functions are addressed, previous solutions led to
significantly high overhead.
Although scalable synthesis methods for large reversible functions are available,
how to derive the respective embeddings is unknown so far.

In this paper, we are addressing this issue by lifting embedding from truth
table based approaches to symbolic ones.
For this purpose, we are exploiting  observations by Bennett on upper bounds and
general embeddings.
Additionally, we use the symbolic representation of BDDs.
However, we first consider the theoretical complexity of the embedding problem.

\section{Bounds for the Number of Additional Lines}
\label{sec:bounds}

\subsection{Upper Bound}
\label{sec:upper-bound}
An upper bound for the number of lines can easily be determined as described by
the following proposition.

\begin{proposition}
  \label{prop:upper-bound}
  Given a function~$f\in\bfunc_{n,m}$.
  Then at most~$\ell(f)=n$ additional lines are required to embed~$f$.
\end{proposition}
\begin{IEEEproof}
  The value of~$\mu(f)$ is maximized if there exists one~$\vec y\in\bbbb^m$
  such that for all~$\vec x\in\bbbb^n$ we have~$f(\vec x)=\vec y$.
  In this case~$\mu(f)=2^n$. Hence, $\ell(f)=\lceil \log_2 2^n\rceil=n$
\end{IEEEproof}

\subsection{Lower Bounds}
\label{sec:minimal-embedding}
In this section, we concern ourselves with the following question: Given a Boolean
function $f\in\bfunc_{n,m}$ and some $\ell\geq 0$, does $\ell=\ell(f)$ hold?
We prove two $\coNP$ lower bounds in this section, one even when $m$ is
a fixed and one even when $\ell$ is fixed.

It is clear that we cannot hope for an efficient procedure for computing an
optimal embedding for a given $f\in\bfunc_{n,m}$ if we cannot even compute
$\ell(f)$ efficiently.

\begin{proposition}{\label{P PP}}
For every fixed $m\geq 1$ it is $\coNP$-hard to decide, given $f\in\bfunc_{n,m}$
with $n\geq m$, whether $\ell(f)=n-m$.
\end{proposition}
\begin{IEEEproof}
We give a polynomial time many-one reduction from the validity problem, i.e.
to decide for a given propositional formula whether it is valid, a
 $\coNP$-complete problem.
Let $\varphi$ be a propositional formula over the variables $\{x_1,\ldots,x_j\}$.
We put $n\defeq j+m$.
We will compute in polynomial time
an $m$-tuple $t=(\psi_1,\ldots,\psi_m)$ of propositional formulas over
the variables $\{x_1,\ldots,x_{j+m}\}$ such that
$\varphi$ is valid if, and only if, $\ell(f_t)=n-m=j$.
We put
\begin{equation}
\label{eq:constrP PP}
\psi_i(x_1,\dots,x_{j+m})\quad\defeq\quad x_{j+i}\wedge\varphi(x_1,\dots,x_j)
\end{equation}
for each $i\in\{1,\ldots,m\}$.
The correctness of the reduction follows immediately from the following equivalences:
\begin{eqnarray*}
&&\varphi\text{ is valid } \\
& \Leftrightarrow &
\#\{\vec x\in\bbbb^j\mid \varphi(\vec x)=1\}=2^j\\
& \stackrel{\text{\eqref{eq:constrP PP}}}{\Leftrightarrow} &
\forall \vec y\in f_t(\bbbb^{j+m}): \#\{\vec x\in\bbbb^{j+m}\mid f_t(\vec x)=\vec y\}
=\frac{2^{j+m}}{2^m}\\
& \Leftrightarrow &
\mu(f_t)=2^j\\
& \Leftrightarrow &
\ell(f_t)=j=n-m
\end{eqnarray*}
\end{IEEEproof}

Looking at the proof of Proposition \ref{P PP} we note that we required
$\ell$ to be part of the input for the lower bound to work since we fixed
$m$. Moreover the lower
bound already holds when $m=1$ and thus when $\ell=n-1$.
The dual question arises whether a lower bound can be proven when $\ell$ is
a fixed constant.
Indeed, we prove that computing $\ell(f)$ is already $\mathsf{coNP}$-hard
for every {\em fixed} $\ell\geq 0$ (i.e. the input to the problem
only consists of $f$ and the value $\ell$ is treated as a fixed constant).

\begin{proposition}
\label{prep:injectiveness}
For each fixed $\ell\geq 0$ it is $\mathsf{coNP}$-hard to decide
for a given $f\in\bfunc_{n,m}$ whether $\ell(f)=\ell$.
\end{proposition}

\begin{IEEEproof}
We give a polynomial time many-one reduction from the
validity problem for propositional formulas.
For simplicity we only give the proof for $\ell=0$.
The case $\ell>0$ works completely analogously.
Fix a propositional formula~$\varphi\in\bfunc_n$ over the
variables~$\{x_1,\dots x_n\}$.
We will compute in polynomial time a function $f=(f_1,\ldots,f_n)$, where
each $f_i$ is presented by a propositional formula
$\varphi_i$ over the variables $\{x_1,\ldots,x_n\}$ such that~$\varphi$
is valid if, and only if,~$f$ is injective.
We set
\begin{equation}
  \label{eq:reduction}
  \varphi_i(x_1,\dots,x_n)\defeq x_i\land \varphi(x_1,\dots,x_n)\land \varphi(0,\dots,0)
\end{equation}
for each~$i\in\{1,\dots,n\}$.

For correctness, we have to show that $\varphi$ is valid if, and only
if, $\ell(f)=0$, which in turn is equivalent to $f$ being injective.

Let us first assume that~$\varphi$ is valid.
Then $\varphi_i$ is logically equivalent to $x_i$ for each $i\in\{1,\ldots,n\}$
by definition~\eqref{eq:reduction}.
Thus, $f$ is equivalent to the identity~$\mathrm{id}_n\in\bfunc_{n,n}$, i.e.
$\mathrm{id}_n(\vec x)\defeq\vec x$ for each~$\vec x\in\bbbb^n$.

Conversely, assume that~$\varphi$ is not valid.
Then there exists some~$\vec x\in\bbbb^n$ with~$\varphi(\vec x)=0$.
We make a case distinction.
In case~$\vec x=0^n$, we immediately have that~$\varphi_i$ is logically equivalent to
$0$ and thus~$f$ is not
injective.
Now assume that~$\vec x\neq 0^n$.
Since~$\varphi(0^n)=0^n$ and~$\varphi(\vec x)=0^n$ and also~$\vec x\neq 0^n$, the function
$f$  is not injective.
\end{IEEEproof}

We note that two simple lower bound corollaries follow from Proposition \ref{P PP} and
\ref{prep:injectiveness}:
(1) it is \coNP-hard to decide if for a given function~$f\in\bfunc_{n,n}$ and a
given~$\ell\geq 0$ whether there exists an injective~$g\in\bfunc_{n,n+\ell}$ that
embeds~$f$,
(2) there exists no polynomial time algorithm that computes, given a function
$f\in\bfunc_{n,n}$, the minimal~$\ell\geq0$ such that~$f$ can be embedded by some
function~$g\in\bfunc_{n,n+\ell}$ unless~$\mathsf{P}=\mathsf{NP}$.

\section{Determining the Number of Additional Lines}
\label{sec:calculating-lines}
In this section we propose three algorithms to determine the number of
additional lines.  The first algorithm approximates the number of lines while
the second and third one determine the minimal number.

\subsection{Heuristic Cube-based Approach}
\label{sec:heuristic-cube-based}
The first approach approximates the number of cubes and is based on the PLA
representation of a multiple output function.  The approach is an extension of
an algorithm presented
in~\cite{conf/date/WilleKD11}\footnote{In~\cite{conf/date/WilleKD11}, the
  algorithm does not consider the OFF-set of the function and, hence, may return
  an under approximation.}.  Given a function, one can approximate the minimal
number of additional variables that are needed to embed the function using the
following algorithm.

\begin{figure}[t]
  \centering
  \def\tabcolsep{2pt}
  \newcolumntype{C}{>{$}c<{$}}
  \newcommand{\numberwithanchor}[2]{\tikz[remember picture,baseline=(#1.base)] \node[font=\itshape,inner sep=0pt,outer sep=0pt] (#1) {#2};}
  \begin{tabular}{CCCCC|CCC>{\quad\it}c}
    x_1 & x_2 & x_3 & x_4 & x_5 & y_1 & y_2 & y_3 &   \\\cline{1-8}
    1   & -   & -   & 0   & -   & 1   & 0   & 0   & 8 \\
    0   & 0   & -   & -   & -   & 0   & 1   & 0   & 8 \\
    1   & 1   & -   & -   & 1   & 0   & 0   & 1   & \numberwithanchor{a}{4} \\
    -   & 1   & 0   & -   & -   & 0   & 0   & 1   & \numberwithanchor{b}{8} \\
    1   & 0   & -   & 1   & -   & 1   & 0   & 1   & \numberwithanchor{c}{4} \\
    1   & 1   & -   & 1   & 0   & 1   & 0   & 1   & \numberwithanchor{d}{2} \\
  \end{tabular}
  \begin{tikzpicture}[overlay,remember picture]
    \coordinate (m) at ([xshift=20pt] $(a)!.5!(b)$);
    \draw[thin] ([xshift=2pt] a.east) -- (m) node[right,font=\itshape] {12} -- ([xshift=2pt] b.east);
    \coordinate (m) at ([xshift=20pt] $(c)!.5!(d)$);
    \draw[thin] ([xshift=2pt] c.east) -- (m) node[right,font=\itshape] {6} -- ([xshift=2pt] d.east);
  \end{tikzpicture}
  \caption{PLA representation of a Boolean function}
  \label{fig:example_heur_alg}
\end{figure}

\algbegin Algorithm H (Heuristic Cube-based Approach). This algorithm
approximates $\mu(f)$ by~$\hat\mu(f)$ for a given
function~$f\in\mathcal{B}_{n,m}$ given in PLA
representation~$P_f:C\mapsto\mathcal{P}(f)$.  An auxiliary array~\codemu{o}
for~$o\in\mathcal{P}(f)$ is used to compute possible candidates
for~$\hat\mu(f)$.

\algstep H1. [Initialize.] Set~$\codemu{o}\leftarrow 0$ for
each~$o\in\mathcal{P}(f)$.

\algstep H2. [Loop over $C$ and update~$\code{MU}$.] For each~$c\in C$,
set~$\codemu{P_f(c)}\leftarrow\codemu{P_f(c)}+\#\onset(c)$.

\algstep H3. [Count~$\offset(f)$ in~$\code{MU}$.]
Set~$\codemu{\emptyset}\leftarrow\#\offset(f)$.

\algstep H4. [Determine~$\hat\mu(f)$.]
Set
\[\hat\mu(f)\leftarrow\max\{\codemu{o}\mid o\in\mathcal{P}(f)\}.\qquad\slug\]

\begin{remark}
  Although the set~$\mathcal{P}(f)$ is exponentially large, Algorithm~H can be
  efficiently implemented, since it only needs to consider those elements that
  are in the image of~$P_f$.  Step 3 can be implemented using BDDs.
\end{remark}

\medskip
\begin{example}\label{exa:expl_two_level_descr}
  Consider the function given in Fig.~\ref{fig:example_heur_alg}.  Algorithm~H
  basically assigns the number of input patterns that are represented by an
  input cube to each line of the PLA representation.  In order to approximate,
  we assume that the $0$s in the table are part of the output pattern.  Values
  of lines with the same output pattern are added.  The OFF-set of the function
  is not mentioned in the table representation and can be described by the
  cube~\texttt{011-{}-} that corresponds to 4 input pattern.  The algorithm
  computes, that approximately~$4+8=12$ input patterns map to the output
  pattern~\texttt{001} which corresponds to the computed approximation
  for~$\mu(f)$ obtained from Algorithm~H.  Hence, $3+\lceil\log_2 12\rceil = 7$
  lines may be sufficient to realize this function as a reversible circuit.
\end{example}

\medskip
\noindent
The determined value for~$\hat\mu(f)$ is still an approximation, since overlaps
of the input cubes are not yet considered.  For example, the two input cubes
discussed in Example~\ref{exa:expl_two_level_descr} share some equal input
patterns, i.e.~the determined number of~$12$ occurrences for the output
pattern~\texttt{001} is an approximation.  In fact the output patterns that are
assumed for the approximation may have nothing in common with the \emph{real}
output patterns of the function that is described by the PLA representation.
Consequently, $\hat\mu(f)$ can be smaller than $\mu(f)$.  An example for this
case is given in the following section.  However, the results from the experimental
evaluation presented in Section~\ref{sec:exper-eval} will show that the
approximation is often close to the exact value.

\subsection{Exact Cube-based Approach}
\label{sec:extending}
If we had a PLA representation~$P_f':C'\to\mathcal{P}(f)$ in which no
input cube overlaps, i.e.~$\onset(c_i)\cap\onset(c_j)=\emptyset$ for all
pairwise~$c_i,c_j\in C'$, one~\emph{can} guarantee that~$\hat\mu(f)=\mu(f)$
after applying Algorithm~H to~$P_f'$.  The expressions represented by such PLAs
are called \emph{disjoint sum-of-products}~(DSOP) in the literature and several
algorithms to derive such a representation have been described in the past.  The
most recent results can be found e.g.~in~\cite{BCLP:13,BE:10}.  We create a DSOP
representation based on an algorithm described in~\cite{conf/date/WilleKD11}
that particularly addresses multiple-output functions.

Please note that the compact representation of a PLA representation highly
depends on overlapping input cubes.  As a result, the PLA representation of the
DSOP expression is possibly very large.  In the worst case, the whole truth
table is reconstructed.

\algbegin Algorithm D (Disjoint Sum-of-Product Computation). Given a PLA
representation~$P_f:C\to\mathcal{P}(f)$ of a function $f\in\bfunc_{n,m}$, this
algorithm computes a new PLA representation~$P_f':C'\to\mathcal{P}(f)$ where no
input cubes overlap, i.e.~$\onset(c_i)\cap\onset(c_j)=\emptyset$ for all
pairwise different~$c_i,c_j\in C'$.  In the algorithm, $P_f$ and~$P_f'$ are
treated as mutable relations.

\algstep D1. [Terminate?] If~$P_f=\emptyset$, terminate.

\algstep D2. [Pick an entry from~$P_f$ and iterate over~$P_f'$.]  Pick and
remove one entry $e=(c,o)$ from~$P_f$, i.e.~set~$P_f\leftarrow
P_f\setminus\{e\}$.  If there exists an overlapping cube~$e'=(c',o')\in P'_f$,
i.e.
\[ c\land c'\neq\bot, \]
perform step~3.  If no such~$e'$ exists, set~$P_f'\leftarrow
P_f'\cup\{e\}$ and return to step~1.

\algstep D3. [Update~$P_f$ and~$P_f'$.] Remove the entry~$e'$ from~$P'_f$,
i.e.~set~$P'_f\leftarrow P'_f\setminus\{e'\}$.  Keep one ``remaining part''
for~$P_f$, i.e.
\[ P_f\leftarrow P_f\cup\{(c\land\bar c',o)\} \]
and the other one for~$P'_f$, i.e.
\[ P'_f\leftarrow P'_f\cup\{(c'\land\bar c,o')\}. \]
Add the intersection to~$P'_f$ by combining the output sets, i.e.
\[ P'_f\leftarrow P'_f\cup\{(c\land c', o\cup o')\}.\qquad\slug \]

\begin{figure}[t]
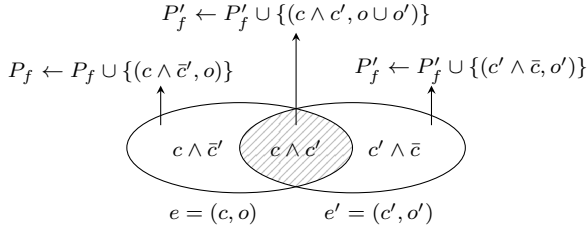

  \centering
  \tikzUnfoldingSubsets
  \caption{Illustration for step~3 in Algorithm~D}
  \label{fig:step-3-explanation}
\end{figure}

\medskip\noindent Algorithm~D transforms the initial PLA representation $P_f$
into the initial empty PLA representation~$P'_f$ which represents the same
function as DSOP expression.  As long~$P_f$ is non-empty an entry~$e=(c,o)$ is
chosen for which the following case distinction is applied.  If~$c$ does not
intersect with any other input cube in~$P'_f$, the entry~$e$ is removed
from~$P_f$ and directly added to~$P'_f$.  Otherwise, i.e.~if there exists an
entry~$e'=(c',o')$ with~$c\land c'\neq\bot$, step~3 is performed.  What happens
in this step after~$e'$ has been removed from~$P_f'$ is best illustrated by
means of Fig.~\ref{fig:step-3-explanation}.  The ON-set of~$c\lor c'$ is
partitioned into three parts.  The part of~$c$ that does not intersect~$c'$
remains in~$P_f$, and analogously, the part of~$c'$ that does not intersect~$c$
remains in~$P_f'$.  The intersection is also added to~$P_f'$, however, the
corresponding output functions are combined.

\begin{figure}[t]
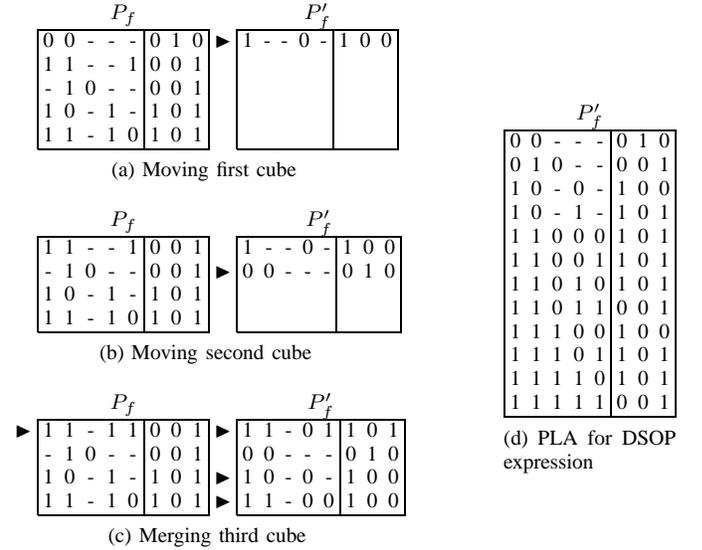

  \footnotesize
  \def\m{$\blacktriangleright$}
  \def\tabcolsep{2pt}
  \begin{minipage}{.6\linewidth}
  \subfloat[Moving first cube]{%
  \begin{tabular}[b]{c|ccccc|ccc|c|ccccc|ccc|}
    \multicolumn{1}{c}{} & \multicolumn{8}{c}{$P_f$} &
    \multicolumn{1}{c}{} & \multicolumn{8}{c}{$P'_f$} \\
    \cline{2-9}\cline{11-18}
    \phantom{\m}&0&0&-&-&- & 0&1&0 &\m& 1&-&-&0&- & 1&0&0 \\
    &1&1&-&-&1 & 0&0&1 &  &  & & & &  &  & &  \\
    &-&1&0&-&- & 0&0&1 &  &  & & & &  &  & &  \\
    &1&0&-&1&- & 1&0&1 &  &  & & & &  &  & &  \\
    &1&1&-&1&0 & 1&0&1 &  &  & & & &  &  & &  \\\cline{2-9}\cline{11-18}
  \end{tabular}}

  \subfloat[Moving second cube]{%
  \begin{tabular}[b]{c|ccccc|ccc|c|ccccc|ccc|}
    \multicolumn{1}{c}{} & \multicolumn{8}{c}{$P_f$} &
    \multicolumn{1}{c}{} & \multicolumn{8}{c}{$P'_f$} \\
    \cline{2-9}\cline{11-18}
    \phantom{\m}&1&1&-&-&1 & 0&0&1 &  & 1&-&-&0&- & 1&0&0 \\
    &-&1&0&-&- & 0&0&1 &\m& 0&0&-&-&- & 0&1&0 \\
    &1&0&-&1&- & 1&0&1 &  &  & & & &  &  & &  \\
    &1&1&-&1&0 & 1&0&1 &  &  & & & &  &  & &  \\\cline{2-9}\cline{11-18}
  \end{tabular}}

  \subfloat[Merging third cube]{%
  \begin{tabular}[b]{c|ccccc|ccc|c|ccccc|ccc|}
    \multicolumn{1}{c}{} & \multicolumn{8}{c}{$P_f$} &
    \multicolumn{1}{c}{} & \multicolumn{8}{c}{$P'_f$} \\
    \cline{2-9}\cline{11-18}
  \m&1&1&-&1&1 & 0&0&1 &\m& 1&1&-&0&1 & 1&0&1 \\
    &-&1&0&-&- & 0&0&1 &  & 0&0&-&-&- & 0&1&0 \\
    &1&0&-&1&- & 1&0&1 &\m& 1&0&-&0&- & 1&0&0 \\
    &1&1&-&1&0 & 1&0&1 &\m& 1&1&-&0&0 & 1&0&0 \\\cline{2-9}\cline{11-18}
  \end{tabular}}
  \end{minipage}
  \hfill
  \begin{minipage}{.3\linewidth}
  \hfill
  \subfloat[PLA for DSOP expression]{%
  \begin{tabular}[b]{|ccccc|ccc|}
    \multicolumn{8}{c}{$P'_f$} \\
    \hline
    0&0&-&-&-& 0&1&0 \\
    0&1&0&-&-& 0&0&1 \\
    1&0&-&0&-& 1&0&0 \\
    1&0&-&1&-& 1&0&1 \\
    1&1&0&0&0& 1&0&1 \\
    1&1&0&0&1& 1&0&1 \\
    1&1&0&1&0& 1&0&1 \\
    1&1&0&1&1& 0&0&1 \\
    1&1&1&0&0& 1&0&0 \\
    1&1&1&0&1& 1&0&1 \\
    1&1&1&1&0& 1&0&1 \\
    1&1&1&1&1& 0&0&1 \\\hline
  \end{tabular}}
  \end{minipage}
  \caption{DSOP computation for the function in Fig.~\ref{fig:example_heur_alg}}
  \label{fig:unfolding-example}
\end{figure}

\begin{example}
  \label{ex:extending}
  An example application of Algorithm~D is demonstrated in
  Fig.~\ref{fig:unfolding-example} based on the PLA of Fig.~\ref{fig:pla}.
  Clearly, the first cube can be moved without merging since~$P_f'$ is initially
  empty.  Affected cubes in steps are marked by~`$\blacktriangleright$'.  Also
  the second cube can be moved without merging since it does not intersect the
  first one.  The third cube to be moved from~$P_f$ to~$P_f'$ has the input
  pattern~$c=x_1x_2x_5$.  It intersects with the current first cube in~$P_f'$
  which has the input pattern~$c'=x_1\bar x_4$.  The ``remaining part''
  for~$P_f$ is
  \[ x_1x_2x_5\land\overline{x_1\bar x_4}=x_1x_2x_4x_5. \]
  Analogously we have
  \[ \overline{x_1x_2x_5}\land x_1\bar x_4=x_1\bar x_2\bar x_4\lor x_1x_2\bar
  x_4\bar x_5 \]
  which yields two new entries to be added to~$P_f'$.  The current entry is
  updated by the intersection
  \[ x_1x_2x_5 \land x_1\bar x_4=x_1x_2\bar x_4x_5 \]
  for which the corresponding output functions are merged.  All other steps are
  not shown explicitly.  The final PLA representation for~$P_f'$ is given in
  Fig.~\ref{fig:unfolding-example}(d).
\end{example}

\subsubsection*{Correctness and completeness} We now prove that Algorithm~D is
sound and complete.
\begin{lemma}
Algorithm~D is complete.
\end{lemma}
\begin{IEEEproof}
  We show that the algorithm terminates for every input function~$f$.  For this
  purpose, we show that in every iteration the size of the ON-set of the function
  represented by~$P_f$ decreases.  Consequently, we eventually
  have~$P_f=\emptyset$ which leads to termination in step~1.  We perform a case
  distinction on whether a cube~$e'$ exists in step~2.

  If such a cube~$e'$ does not exist, $e=(c,o)$ is removed from~$P_f$ and we
  have~$c\neq\bot$.  Further, nothing is added to~$P_f$ in this case.

  Otherwise, first~$c$ is removed from the input cubes and
  afterwards~$c\land\bar c'$ is added.  However, since by assumption~$c\land
  c'\neq\bot$, we have~$\#\onset(c\land\bar c')<\#\onset(c)$.
\end{IEEEproof}

\begin{lemma}
  Algorithm~D is sound.
\end{lemma}
\begin{IEEEproof}
  Clearly, $P'_f$ does not contain overlapping input cubes due to step~3.  We
  now show that in step~1 the function represented by~$P_f\cup P'_f$ equals~$f$.
  Since eventually~$P_f=\emptyset$, $P'_f$ represents~$f$.  We perform a case
  distinction on whether a cube~$e'$ exists in step~2.

  If such a cube~$e'$ does not exist, the case is trivial, since~$e$ is removed
  from~$P_f$ and directly added to~$P'_f$.  For the case that a cube~$e'$ does
  exist one can readily observe based on the partition in
  Fig.~\ref{fig:step-3-explanation} that the ON-set of~$P_f\cup P'_f$ does not
  change.  Since the output functions are combined for the intersection~$c\land
  c'$, also the functional semantics of~$f$ is preserved.
\end{IEEEproof}

\begin{figure}[t]
  \footnotesize
  \def\tabcolsep{2pt}
  \centering
  \begin{tabular}[b]{|ccccc|ccc|}
    \multicolumn{8}{c}{$P'_f$} \\
    \hline
    0&0&-&-&-& 0&1&0 \\
    0&1&0&-&-& 0&0&1 \\
    1&0&-&0&-& 1&0&0 \\
    1&0&-&1&-& 1&0&1 \\
    1&1&-&1&1& 0&0&1 \\
    1&1&0&0&-& 1&0&1 \\
    1&1&0&1&0& 1&0&1 \\
    1&1&1&0&0& 1&0&0 \\
    1&1&1&0&1& 1&0&1 \\
    1&1&1&1&0& 1&0&1 \\\hline
  \end{tabular}
  \caption{PLA representation after post compaction}
  \label{fig:post-compaction}
\end{figure}

\subsubsection*{Post compaction} It is possible to compact the resulting PLA
representation that is returned by Algorithm~D.  For this purpose we create a
BDD for each occurring output pattern from the input cubes.  This step is only
sound because the input cubes do not overlap.  By traversing all paths in the
BDD we can obtain a new PLA representation for each output cube.  It turns out
that the PLA representation for the input cubes is more compact compared to the
one that resulted from applying Algorithm~D.

\begin{example}
  The application of Algorithm~D in Example~\ref{ex:extending} yields a PLA
  representation that consists of 12 cubes.  Applying post compaction as post
  process yields a PLA representation that consists of 10 cubes
  (cf.~Fig.~\ref{fig:post-compaction}).  In the experimental evaluation we will
  demonstrate that much higher compression can be achieved with this method.
\end{example}

\subsubsection*{An under approximating example} As discussed in
Section~\ref{sec:heuristic-cube-based}, Algorithm~H can yield an
under-approximation for~$\mu(f)$.  Consider the following PLA representation for
a function with 5 input variables, 3 output variables, and 6 monoms:
\[
  \def\tabcolsep{2pt}
  \begin{tabular}[b]{|ccccc|ccc|}
    \hline
0&0&-&1&-& 0&0&1 \\
0&0&0&1&0& 0&0&1 \\
1&1&1&1&-& 0&0&1 \\
-&-&-&1&-& 0&1&0 \\
1&-&-&-&1& 0&1&1 \\
1&-&-&-&0& 0&1&1 \\\hline
  \end{tabular}
\]
Algorithm~H will compute the following output pattern occurrences:
\[ 000\mapsto8 \qquad 001\mapsto7 \qquad 010\mapsto16 \qquad 011\mapsto16 \]
Note that the sum of all occurrences is 47 although the function only has 32
output patterns.  Algorithm~H yields~$\hat\mu(f)=16$ which corresponds to $4$
additional output variables and therefore a total of~$7$ variables in a
reversible embedding.

After applying Algorithm~D (together with post compaction) one obtains the
following equal PLA representation with no overlapping input cubes:
\[
  \def\tabcolsep{2pt}
  \begin{tabular}[b]{|ccccc|ccc|}
    \hline
    0&1&-&1&-& 0&1&0 \\
    0&0&-&1&-& 0&1&1 \\
    1&-&-&-&-& 0&1&1 \\\hline
  \end{tabular}
\]
For this PLA representation Algorithm~H will compute the following output
pattern occurrences:
\[ 000\mapsto8 \qquad 010\mapsto4 \qquad 011\mapsto20 \] Since the PLA is
representing a DSOP expression we have~$\hat\mu(f)=\mu(f)=20$ which corresponds
to~$5$ additional output variables and a total of~$8$ variables in a reversible
embedding.

\subsection{BDD-based Approach}
\label{sec:bdd-lines}
All approaches presented thus far use a PLA representation in 
computing or approximating~$\mu(f)$.  However, sometimes such a
representation is not available but the functions that are being considered can
be represented as BDDs.  In this section an algorithm is described that
computes~$\mu(f)$ directly on the BDD representation of an irreversible
function~$f\in\bfunc_{n,m}$ in memory with~$f(x_1,\dots,x_n)=(y_1,\dots,y_m)$.
For this purpose, first the characteristic function~$\chi_f$ is computed as
described in~\eqref{eq:characteristic}.

For this purpose, the BDD of the characteristic function is constructed
assuming the variable order $x_1<\dots<x_n<y_1<\dots<y_m$.
Then, let~$V_x$ be the set of all vertices that are labeled~$x_i$ for some~$i$ and
whose immediate parent is labeled~$y_j$ for some~$j$. The on-sets of
the functions represented by each vertex in~$V_x$ form a partition of all $2^n$
input patterns. This can be exploited by using the following proposition.

\begin{proposition}
  In the BDD of~$\chi_f$ every path from the start vertex to a vertex in $V_x$
  visits all variables~$y_1,\dots,y_m$.  Further, each vertex in~$V_x$ has only
  one incoming edge.
\end{proposition}

\begin{IEEEproof}
  Assume that there is a path from the start vertex to a vertex~$v$ in~$V_x$ in
  which a variable~$v_j$ is not visited.  Then, the monom represented by~$v$
  maps to more than one output in~$f$ which contradicts that~$f$ is a function.
  The same argument holds when we assume that $v$ has more than one incoming
  edge.
\end{IEEEproof}

Now~$\mu(f)$ can easily be computed using the vertices in~$V_x$, i.e.:
\begin{equation}
  \label{eq:muf-with-char-bdd}
  \mu(f)=\max\{\#\onset(\sigma(v))\mid v\in V_x\}
\end{equation}

\begin{figure}[t]
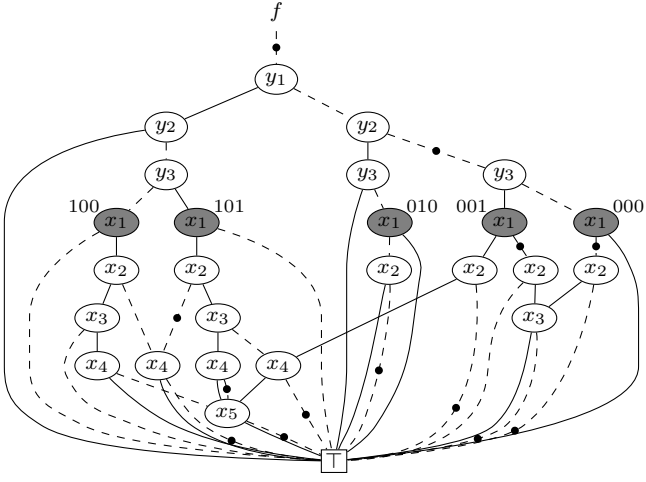

  \centering
  \tikzCharBDDExampleWithZeros
  \caption{BDD for the characteristic function of the example in
    Fig.~\ref{fig:example_heur_alg}}
  \label{fig:char-bdd}
\end{figure}

\begin{example}
  The BDD for the characteristic function of the example in
  Fig.~\ref{fig:example_heur_alg} is given in Fig.~\ref{fig:char-bdd}.  Vertices
  from the set $V_x$ are filled in gray and annotated by the output pattern they
  are mapping to.  Based on the paths one can count the minterms of each node
  which results in:
  \[ 100\mapsto4 \quad 101\mapsto9 \quad 010\mapsto8 \quad 001\mapsto6 \quad
  000\mapsto5 \] The numbers coincide with the results from
  Example~\ref{ex:extending}.
\end{example}

\section{Embedding Irreversible Functions}
\label{sec:embedding}

In this section, we describe two approaches that construct a reversible
embedding for a given irreversible function~$f$.  The first approach takes~$f$
as a DSOP expression and guarantees a minimal number of additional lines.  The
second approach takes~$f$ as a binary decision diagram and is heuristic.

\subsection{Exact Cube-based Approach}
\label{sec:exact-cube-based}
First, we will describe an exact embedding approach with respect to the number
of additional lines that makes use of Algorithms~H and~D from the previous
section.  Given an irreversible function represented as a PLA, it is first
transformed to represent a DSOP expression to determine the optimal number of
additional lines.  Then a reversible function is created by traversing all cubes
of the DSOP expression.  The algorithm requires the function to be represented
as a DSOP expression in order to guarantee that no two input cubes have a
non-empty intersection.  Also, the algorithm creates a partial reversible
function, i.e.~not for all input patterns an output pattern is specified.
However, all specified patterns in the reversible function are fully specified,
i.e.~they have no \emph{don't-care} values.

We are making use of two helper functions in the following algorithm which are
defined as follows.

Given a function~$f\in\bfunc_{n,m}$ and a set of output
functions~$o\in\mathcal{P}(f)$, the function
\[
  \cube(o)\defeq\bigwedge_{i=1}^m
  \begin{cases}
    y_i & \text{if $f_i\in o$,} \\
    \bar y_i & \text{otherwise,}
  \end{cases}
\]
creates a cube that contains a positive literal~$y_i$ if the function~$f_i$ is
contained in~$o$ and a negative literal~$\bar y_i$ otherwise.

Given a set of variables~$x_1,\dots,x_n$ the function
\[ \inc(x_1,\dots,x_n)=(s_1,\dots,s_n) \] computes the increment of the integer
representation given by~$x_nx_{n-1}\dots x_1$, i.e.
\[ s_i \defeq x_i \oplus \bigwedge_{j=1}^{i-1}x_j \qquad\text{for
  $i\in\{1,\dots,n\}$}. \]

Note that both functions, $\cube$ and~$\inc$, are easy to implement using BDD
manipulation.  We are now ready to formulate the algorithm to embed a truth
table based on a function's PLA representation.

\algbegin Algorithm E (Cube-based Embedding). Given a
function~$f\in\bfunc_{n,m}$ represented as a DSOP, this algorithm
generates a partial reversible function~$g\in\bfunc_{r,r}$ with
\[g(\constant_1,\dots,\constant_p,x_1,\dots,x_n)
=(y_1,\dots,y_m,\gamma_1,\dots,\gamma_{\ell(f)})\] and $r=p+n=m+\ell(f)$ that
embeds~$f$ and is represented as its characteristic~$\chi_g$ function using a
BDD.  Let~$P_f=\{(c_1,o_1),\dots,(c_k,o_k)\}$ be the PLA representation for the
DSOP expression of~$f$, i.e.~no input cubes overlap.

\algstep E1. [Initialization.] Let~$\chi_g\leftarrow\bot$ be a BDD with variable
ordering
\[\constant_1<y_1<\constant_2<y_2<\dots<x_n<\gamma_{\ell(f)},\]
i.e.~inputs and outputs of~$\chi_g$ appear in alternating order.
Set~$\ell\leftarrow\ell(f)$ where~$\ell(f)=\lceil\log_2\mu(f)\rceil$ is obtained
from Algorithm~H.  Further, set~$j\leftarrow0$ and~$\codecnt{o}\leftarrow0$
for~$o\in\mathcal{P}(f)$.

\algstep E2. [Loop over~$C$.] If~$j=k$, terminate.  Otherwise, set~$j\leftarrow
j+1$,~$c\leftarrow c_j$, and~$o\leftarrow o_j$.

\algstep E3. [Add cube to~$\chi_g$.]
Let
\[(s_1,\dots,s_{\ell})=\inc^q(\gamma_1,\dots,\gamma_{\ell})\]
where~$q=\codecnt{o}$, i.e.,~$\inc$ is applied~$q$ times to the garbage
outputs~$\gamma_1,\dots,\gamma_{\ell}$ and the values are stored
in~$s_1,\dots,s_{\ell}$.  Also, let
\[\{d_1,\dots,d_t\}=\dc(c)\qquad\text{with~$d_1<d_2<\dots< d_t$}\] be the
indexes of variables which are set~\emph{don't-care} in the input cube~$c$.  Let
\begin{equation}
  \label{eq:heuristic-embedding-add-cubes}
    e=\underbrace{c}_{\text{$x$'s}}
    \land\underbrace{\cube(o)}_{\text{$y$'s}}
    \land\underbrace{\bigwedge_{i=1}^p\bar \kappa_i}_{\text{$\kappa$'s}}
    \land\underbrace{\bigwedge_{i=1}^t\left(x_{d_i}\leftrightarrow s_i\right)
    \land\bigwedge_{i=t+1}^{\ell}\bar s_i}_{\text{$\gamma$'s}}
\end{equation}
and set~$\chi_g\leftarrow\chi_g\lor e$.  Also,
set~$\codecnt{o}\leftarrow\codecnt{o}+\#\onset(c)$.  Return to
step~2.\qquad\slug

\medskip\noindent In step~1, an empty BDD~$\chi_g$ is created that interleaves
inputs and outputs in its variable ordering.  The size of the BDD is determined
by calculating the minimal number of additional lines~$\ell(f)$ using
Algorithm~H after applying Algorithm~D.  The auxiliary array~$\code{CNT}$ is
used to store how often an output pattern~$o$ has been used and is initially
initialized to~$0$.

Step~2 manages the algorithm's loop over step~3.  In each iteration one entry
of~$P_f$ is added to~$g$.  Step~3 creates a cube~$e$
in~\eqref{eq:heuristic-embedding-add-cubes} for the characteristic
function~$\chi_g$ based on the entry~$(c,o)$ of the PLA representation~$P_f$.
This cube~$e$ contains all required ``ingredients,'' i.e.~values for $x$'s,
$y$'s, $\kappa$'s, and~$\gamma$'s referring to inputs, outputs, constants, and
garbage outputs, respectively.
  \begin{itemize}
  \item Input assignments for~$x_1,\dots,x_n$ are directly obtained from the
    input cube~$c$.
  \item Output assignments for~$y_1,\dots,y_m$ are obtained from~$\cube(o)$.
  \item Constants~$\kappa_1,\dots,\kappa_p$ are all assigned~$0$.
  \item The idea is to relate the values of the \emph{don't-care} variables
    of~$c$ to the garbage outputs~$\gamma_1,\dots,\gamma_{\ell}$.  Since there
    may be equal output patterns, an offset~$q=\codecnt{o}$ is taken into
    account.  To calculate the offset, the~`$\inc$' function is applied to
    the~$\gamma$ variables~$q$ times.  If there are less \emph{don't-care}
    variables than garbage outputs, the remaining~$s$ variables are inverted.
    Since~$\ell$ is also obtained based on the number of \emph{don't-care}
    variables in~$c$, there are never more \emph{don't-care} variables than
    garbage outputs in this step, i.e.~$t\le\ell$.
  \end{itemize}
  
\begin{example}
  We apply Algorithm~E to the function with the PLA representation
  \[
  \def\tabcolsep{2pt}
  \begin{tabular}[b]{|ccccc|ccc|}
    \hline
    0&1&-&1&-& 0&1&0 \\
    0&0&-&1&-& 0&1&1 \\
    1&-&-&-&-& 0&1&1 \\\hline
  \end{tabular}
  \]
  that has already been used in Section~\ref{sec:extending}.  Initially we
  set~$\chi_g\leftarrow\bot$. Also we assign~$\codecnt{\{f_2\}}\leftarrow 0$ for
  the first pattern and~$\codecnt{\{f_2,f_3\}}\leftarrow 0$ for the second and
  third pattern.

  Since~$\mu(f)=20$ we have~$\ell(f)=5$ and therefore the reversible
  function~$g\in\bfunc_{8,8}$ maps inputs
  $(\kappa_1,\kappa_2,\kappa_3,x_1,x_2,x_3,x_4,x_5)$ to outputs
  $(y_1,y_2,y_3,\gamma_1,\gamma_2,\gamma_3,\gamma_4,\gamma_5)$.  For the first
  cube we have~$(s_1,\dots,s_5)=(\gamma_1,\dots,\gamma_5)$ and hence we have
  \[
  e_1=
  \bar\kappa_1\bar\kappa_2\bar\kappa_3 \; \bar x_1x_2x_4 \;
  \bar y_1y_2\bar y_3 \;
  (x_3\leftrightarrow \gamma_1)(x_5\leftrightarrow \gamma_2)
  \bar\gamma_3\bar\gamma_4\bar\gamma_5
  \]
  and we set~$\codecnt{\{f_2\}}\leftarrow 4$.  For the second cube we also
  have~$(s_1,\dots,s_5)=(\gamma_1,\dots,\gamma_5)$ and hence we have
  \[
  e_2=
  \bar\kappa_1\bar\kappa_2\bar\kappa_3 \; \bar x_1\bar x_2x_4 \;
  \bar y_1y_2y_3 \;
  (x_3\leftrightarrow \gamma_1)(x_5\leftrightarrow \gamma_2)
  \bar\gamma_3\bar\gamma_4\bar\gamma_5
  \]
  and we set~$\codecnt{\{f_2,f_3\}}\leftarrow 4$.  For the third cube we have
  $(s_1,\dots,s_5)=\inc^4(\gamma_1,\dots,\gamma_5)$, i.e.
  \begin{eqnarray*}
    s_1&=&\gamma_1 \\
    s_2&=&\gamma_2 \\
    s_3&=&\gamma_3\oplus1 \\
    s_4&=&\gamma_4\oplus\gamma_3 \\
    s_5&=&\gamma_5\oplus\gamma_3\gamma_4.
  \end{eqnarray*}
  Hence, we have
  \begin{eqnarray*}
  e_3&=&
  \bar\kappa_1\bar\kappa_2\bar\kappa_3 \; x_1 \;
  \bar y_1y_2y_3 \;
  (x_2\leftrightarrow\gamma_1)(x_3\leftrightarrow\gamma_2) \\
  &\land&
  (x_4\leftrightarrow\bar\gamma_3)
  (x_5\leftrightarrow(\gamma_4\oplus\gamma_3))
  (\bar\gamma_5\oplus\gamma_3\gamma_4)
  \end{eqnarray*}
  and update~$\codecnt{\{f_2,f_3\}}\leftarrow20$.  Overall, the partial
  reversible function embedding~$f$ is given by~$\chi_g=e_1\lor e_2\lor e_3$.
\end{example}

\subsubsection*{Correctness and completeness}
Since the only loop in Algorithm~E is bound by the number of cubes in the PLA
representation, completeness is readily shown and it is left to show soundness.

\begin{lemma}
  Algorithm~E is sound.
\end{lemma}

\begin{IEEEproof}
  To proof soundness we show that
  \begin{enumerate}
    \renewcommand{\labelenumi}{(\roman{enumi})}
  \item the input patterns are unique,
  \item the output patterns are unique, and
  \item $g$ embeds~$f$.
  \end{enumerate}
  Since the PLA represents a DSOP expression for~$f$, it does not contain
  overlapping input cubes, and (i) holds trivially.  Also (iii) follows
  immediately from~\eqref{eq:heuristic-embedding-add-cubes}.  Only~(ii) requires
  some more thorough argument.  We already motivated above that~$t\le\ell$ in
  step~3.  Also since~$\ell$ is obtained from Algorithm H, one can see
  that~$\codecnt{o}\le\codemu{o}$ is invariant.  And since
  further~$\ell\le\log_2\codemu{o}$, the assigned value for the garbage lines
  cannot ``overlap.''
\end{IEEEproof}

\subsection{Heuristic BDD-based Embedding}
\label{sec:heuristic-bdd-based}
In this section, an approach is presented that embeds a function directly using
BDDs.  That is, the possibly costly way of having an PLA representation is
omitted by directly starting from BDDs.  These BDDs must be stored in memory and
may have been created by any algorithm.

For this purpose, the idea of embedding as proposed by
Bennett~\cite{journals/ibmrd/Bennett73} who has initially proven the upper bound
from Proposition~\ref{prop:upper-bound} is adapted.  In his constructive proof
he already applied an explicit embedding which is known as \emph{Bennett
  Embedding} and given as follows:

\begin{theorem}[Bennett Embedding]
\label{lemma:bennett-embedding}
Each function $f\in\bfunc_{n,m}$ is embedded by the function~$g\in
\bfunc_{m+n,m+n}$ such that
\begin{equation}
  g(\kappa_1,\dots,\kappa_m,x_1,\dots,x_n)\defeq
  (y_1,\dots,y_m,\gamma_1,\dots,\gamma_n)
\end{equation}
with
\begin{equation}
  \label{eq:embedding_func}
  y_i(\kappa_1,\dots,\kappa_m,x_1,\dots,x_n)=\kappa_i\oplus f_i(x_1,\dots,x_n)
\end{equation}
and
\begin{equation}
  \label{eq:embedding_garbage}
  \gamma_i(\kappa_1,\dots,\kappa_m,x_1,\dots,x_n)=x_i.
\end{equation}
\end{theorem}

\begin{IEEEproof}
  The embedding is illustrated in Fig.~\ref{fig:embedding-scheme}.  Assume~$g$
  is not injective, hence there is an output pattern that occurs at least twice.
  In particular, the function values for~$\gamma_i$ must equal and according
  to~\eqref{eq:embedding_garbage} also the respective assignments for
  inputs~$x_i$ must equal.  But if the assignments for the inputs~$x_i$ are the
  same, then the assignments for the inputs~$\kappa_j$ must differ and due
  to~\eqref{eq:embedding_func} also the function values for~$y_i$, contradicting
  our assumption.
\end{IEEEproof}

\begin{figure}[t]
  \centering
  \tikzHeuristicEmbeddingScheme
  \caption{Bennett embedding scheme}
  \label{fig:embedding-scheme}
\end{figure}

Conducting the embedding posed by Theorem~\ref{lemma:bennett-embedding} on a
truth table as illustrated in Fig.~\ref{fig:embedding-scheme} is infeasible for
large Boolean functions.  Hence, we propose to perform this embedding directly
using BDDs making use of the characteristic function.  More precisely, given a
function~$f\in\bfunc_{n,m}$, a characteristic function~$\chi_g\in\bfunc_{2m+2n}$
that represents a function~$g\in \bfunc_{m+n,m+n}$ according to
Theorem~\ref{lemma:bennett-embedding} is computed by
\begin{equation}
  \label{eq:characteristic_computation-bennett}
  \begin{array}{rcl}
    \chi_g(\vec \kappa,\vec x,\vec y,\vec \gamma)&=&\bigwedge\limits_{i=1}^m(y_i
    \leftrightarrow(\kappa_i\oplus f_i(x_1,\dots,x_n))) \\[7pt]
    &\land&\bigwedge\limits_{i=1}^n(\gamma_i\leftrightarrow x_i)
  \end{array}
\end{equation}
with~$\vec\kappa=\kappa_1,\dots,\kappa_m$, $\vec x=x_1,\dots,x_n$, $\vec
y=y_1,\dots,y_m$, and $\vec\gamma=\gamma_1,\dots,\gamma_n$ based
on~\eqref{eq:characteristic_computation}.
As the experiments in the next section show, this enables the determination
of an embedding for much larger functions.

\begin{remark}

If we construct a BDD from this function that follows the variable ordering
\[ \kappa_1 < y_1 < \dots < \kappa_m < y_m < x_1 < \gamma_1 < \dots < x_n <
\gamma_n, \] a graph results that is isomorphic to the QMDDs which are used for
synthesis of large reversible functions in~\cite{DBLP:conf/aspdac/SoekenWHPD12}.
These QMDDs~\cite{DBLP:conf/ismvl/MillerT06} are binary and use only Boolean
values for the edge weights, therefore they represent permutation matrices.  To
illustrate the relations between BDD vertices for input and output variables of
a characteristic function and a QMDD vertex consider Fig.~\ref{fig:bddvsqmdd}.
The edges of a QMDD inherently represent an input output mapping which is
explicitly expressed with a BDD for a characteristic function since it contains
both input and output vertices.  In the following, BDDs that represent
characteristic functions of reversible functions are called RC-BDDs.  In fact,
the algorithm for the QMDD-based synthesis presented
in~\cite{DBLP:conf/aspdac/SoekenWHPD12} can be performed on RC-BDDs instead.

\begin{figure}[t]
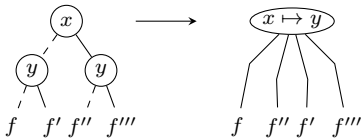

  \centering
  \tikzBDDandQMDD
  \caption{Isomorphism between BDDs and QMDDs}
  \label{fig:bddvsqmdd}
\end{figure}
\begin{figure}[t]
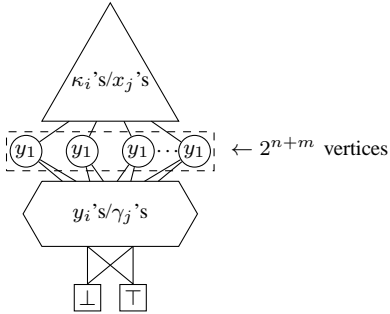

  \centering
  \tikzVariableOrdering
  \caption{Exponential size variable ordering}
  \label{fig:variable-ordering}
\end{figure}

The variable ordering that is interleaving input variables and output variables
is not only necessary in order to directly synthesize the RC-BDD but also
inevitable to keep the number of vertices small.
More precisely, for each RC-BDD there exists two variable orderings which lead
to an exponential number of vertices.
In one of them all input variables are evaluated before all output
variables~(cf.\ Fig.~\ref{fig:variable-ordering}).
Since the RC-BDD represents a reversible function, each input pattern maps to
a distinct output pattern.
Hence, when all input patterns are evaluated first, $2^{m+n}$ vertices to
represent all output patterns remain.
Due to the reversibility, the same applies in case all output patterns are
evaluated before all input patterns.
\end{remark}

\section{Experimental Evaluation}
\label{sec:exper-eval}
We have implemented all algorithms that have been described in
Sections~\ref{sec:calculating-lines} and~\ref{sec:embedding} in C++ using
RevKit~\cite{DBLP:conf/rc/SoekenFWD11}.\footnote{The source code that has been
  used to perform this evaluation is available at www.revkit.org (version 2.0).}
This section presents the results of our evaluation.  Benchmarks were taken from
the LGSynth'93
benchmarks,\footnote{\url{www.cbl.ncsu.edu:16080/benchmarks/lgsynth93/}} from
Dmitri Maslov's benchmarks page,\footnote{\url{www.cs.uvic.ca/~dmaslov/}} and
from RevLib.\footnote{\url{www.revlib.org}} The experimental evaluation has been
carried out on a 3.4~GHz Quad-Core Intel Xeon Processor with 32~GB of main
memory running Linux~3.14.  The timeout for all our experiments was set to
5000~seconds.

\subsection{Determining the Number of Additional Lines}
\begin{table}[t]
  \def\tabcolsep{4pt}
  \caption{Experiments for Determining the Number of Additional Lines}
  \label{tbl:ex-line-determination}
  \begin{tabularx}{\linewidth}{Xrrrrrrrr}
    \hline
    Benchmark & $n$ & $m$ & Bennett & \multicolumn{1}{c}{Heur.~Cube} & \multicolumn{2}{c}{Exact Cube} & \multicolumn{2}{c}{BDD} \\
    \hline
sym9     & 9   & 1   & 10  & 11     & 10  & 0.23    & 10  & 0.01  \\
max46    & 9   & 1   & 10  & \bf10  & 10  & 0.03    & 10  & 0.01  \\
sym10    & 10  & 1   & 11  & \bf11  & 11  & 1.72    & 11  & 0.02  \\
wim      & 4   & 7   & 11  & 10     & 9   & 0.00    & 9   & 0.00  \\
z4       & 7   & 4   & 11  & 12     & 8   & 0.12    & 8   & 0.00  \\
z4ml     & 7   & 4   & 11  & 12     & 8   & 0.06    & 8   & 0.00  \\
sqrt8    & 8   & 4   & 12  & 13     & 9   & 0.09    & 9   & 0.00  \\
rd84     & 8   & 4   & 12  & \bf11  & 11  & 0.17    & 11  & 0.01  \\
root     & 8   & 5   & 13  & 13     & 10  & 0.16    & 10  & 0.00  \\
squar5   & 5   & 8   & 13  & \bf9   & 9   & 0.01    & 9   & 0.00  \\
adr4     & 8   & 5   & 13  & 14     & 9   & 0.17    & 9   & 0.01  \\
dist     & 8   & 5   & 13  & 13     & 10  & 0.25    & 10  & 0.01  \\
clip     & 9   & 5   & 14  & 15     & 11  & 1.08    & 11  & 0.01  \\
cm85a    & 11  & 3   & 14  & 14     & 13  & 0.14    & 13  & 0.00  \\
pm1      & 4   & 10  & 14  & 15     & 13  & 0.01    & 13  & 0.00  \\
sao2     & 10  & 4   & 14  & \bf14  & 14  & 0.14    & 14  & 0.00  \\
misex1   & 8   & 7   & 15  & 15     & 14  & 0.01    & 14  & 0.00  \\
co14     & 14  & 1   & 15  & 19     & 15  & 72.83   & 15  & 0.00  \\
dc2      & 8   & 7   & 15  & 14     & 13  & 0.05    & 13  & 0.00  \\
example2 & 10  & 6   & 16  & 16     & 14  & 1.36    & 14  & 0.01  \\
inc      & 7   & 9   & 16  & \bf14  & 14  & 0.01    & 14  & 0.00  \\
mlp4     & 8   & 8   & 16  & 15     & 13  & 0.19    & 13  & 0.02  \\
ryy6     & 16  & 1   & 17  & 19     & 17  & 157.65  & 17  & 0.00  \\
5xp1     & 7   & 10  & 17  & 17     & 10  & 0.10    & 10  & 0.02  \\
parity   & 16  & 1   & 17  & \bf16  & --- & TO      & 16  & 0.87  \\
t481     & 16  & 1   & 17  & 19     & 17  & 1717.83 & 17  & 0.02  \\
x2       & 10  & 7   & 17  & 19     & 16  & 0.05    & 16  & 0.00  \\
sqr6     & 6   & 12  & 18  & 17     & 12  & 0.05    & 12  & 0.01  \\
dk27     & 9   & 9   & 18  & 16     & 15  & 0.01    & 15  & 0.00  \\
add6     & 12  & 7   & 19  & 20     & 13  & 46.12   & 13  & 0.10  \\
cmb      & 16  & 4   & 20  & 23     & 20  & 7.12    & 20  & 0.01  \\
ex1010   & 10  & 10  & 20  & \bf15  & 15  & 2.89    & 18  & 1.10  \\
C7552    & 5   & 16  & 21  & \bf20  & 20  & 0.01    & 20  & 0.05  \\
decod    & 5   & 16  & 21  & \bf20  & 20  & 0.00    & 20  & 0.05  \\
dk17     & 10  & 11  & 21  & \bf19  & 19  & 0.02    & 19  & 0.01  \\
pcler8   & 16  & 5   & 21  & 23     & 21  & 28.11   & 21  & 0.00  \\
tial     & 14  & 8   & 22  & 23     & 19  & 1007.04 & 19  & 0.18  \\
cm150a   & 21  & 1   & 22  & \bf22  & --- & TO      & 22  & 0.06  \\
alu4     & 14  & 8   & 22  & 24     & 19  & 1270.29 & 19  & 0.13  \\
apla     & 10  & 12  & 22  & \bf22  & 22  & 0.03    & 22  & 0.01  \\
f51m     & 14  & 8   & 22  & 23     & 19  & 556.45  & 19  & 0.20  \\
mux      & 21  & 1   & 22  & \bf22  & --- & TO      & 22  & 0.14  \\
cordic   & 23  & 2   & 25  & 28     & --- & TO      & 25  & 0.06  \\
cu       & 14  & 11  & 25  & 26     & 25  & 0.02    & 25  & 0.00  \\
in0      & 15  & 11  & 26  & \bf25  & 25  & 1.46    & 25  & 0.05  \\
0410184  & 14  & 14  & 28  & \bf14  & 14  & 1227.84 & 14  & 7.40  \\
apex4    & 9   & 19  & 28  & 25     & 26  & 0.95    & 26  & 23.95 \\
misex3   & 14  & 14  & 28  & 30     & 28  & 160.72  & 28  & 17.52 \\
misex3c  & 14  & 14  & 28  & 30     & 21  & 327.16  & 21  & 2.77  \\
cm163a   & 16  & 13  & 29  & 31     & 25  & 625.43  & 25  & 0.01  \\
frg1     & 28  & 3   & 31  & 32     & --- & TO      & 30  & 0.00  \\
bw       & 5   & 28  & 33  & \bf32  & 32  & 0.03    & 32  & 0.04  \\
apex2    & 39  & 3   & 42  & 43     & --- & TO      & 42  & 5.14  \\
pdc      & 16  & 40  & 56  & 61     & 55  & 31.09   & --- & TO    \\
spla     & 16  & 46  & 62  & 65     & 61  & 32.72   & --- & TO    \\
ex5p     & 8   & 63  & 71  & \bf68  & 68  & 0.35    & --- & TO    \\
seq      & 41  & 35  & 76  & 76     & --- & TO      & --- & TO    \\
cps      & 24  & 109 & 133 & 136    & --- & TO      & --- & TO    \\
apex5    & 117 & 88  & 205 & 207    & --- & TO      & --- & TO    \\
e64      & 65  & 65  & 130 & \bf129 & 129 & 0.07    & --- & TO    \\
frg2     & 143 & 139 & 282 & 284    & --- & TO      & --- & TO    \\
\hline
  \end{tabularx}
\end{table}

We have implemented the algorithms from Section~\ref{sec:calculating-lines} in
the RevKit program~`\emph{calculate\_required\_lines}' and evaluated them as
follows.  We have taken the benchmarks in PLA representation and approximated
the number of lines using the heuristic cube-based approach
(Section~\ref{sec:heuristic-cube-based}).  Afterwards we computed the exact
number of additional lines using the exact cube-based approach
(Section~\ref{sec:extending}) and by using the BDD-based approach
(Section~\ref{sec:bdd-lines}).  For the latter one the BDD was created from the
PLA representation.

Table~\ref{tbl:ex-line-determination} list some selected experimental results.
The first three columns list the name of the function together with its number
of inputs and outputs.  The fourth column lists the theoretical upper
bound~(Section~\ref{sec:upper-bound}).  The remaining columns list the number of
lines obtained by the three approaches.  For the two approaches that compute the
number of lines exactly, also the run-time is given.  If no solution has been
found in the given timeout, the cell is labeled with `TO'.  All results for the
heuristic approach have been obtained in a few seconds.  If the approximated
result coincides with the exact one, it is emphasized using bold font.  The
benchmarks are sorted by their theoretical upper bound, i.e.~the sum of the
number of inputs and outputs.

The heuristic cube-based approach is often very close to the exact result.  The
highest measured difference in our experiments was~7 for the function
\emph{add6}.  The function~\emph{apex4} represents the single case in which the
approximated value is smaller than the exact one.

In case of the exact computation the cube-based and BDD-based approaches perform
quite differently.  For the BDD-based approach, the scalability seems to depend
on the size of the function and hence may not scale for functions with more than
50 inputs and outputs.  For some of the larger functions, the cube-based
approach can still obtain a result, however, there are also smaller functions in
which no solution can be found.  This is probably because the scalability of the
approach depends on the number of cubes in the disjoint sum-of-product
representation which does not directly depend on the function size.

\subsection{Cube-based Embedding}
\begin{table}[t]
  \caption{Experiments for Exact Cube-based Embedding}
  \label{tbl:ex-embedding}
  \begin{tabularx}{\linewidth}{Xrrrrr}
    \hline
    Benchmark & $n$ & $m$ & Lines & DSOP Comp. & Run-time \\
    \hline
sym9     & 9  & 1  & 1  & 0.23    & 0.22   \\
max46    & 9  & 1  & 1  & 0.03    & 0.26   \\
sym10    & 10 & 1  & 1  & 1.72    & 0.75   \\
wim      & 4  & 7  & 5  & 0.00    & 0.00   \\
z4       & 7  & 4  & 1  & 0.12    & 0.01   \\
z4ml     & 7  & 4  & 1  & 0.06    & 0.01   \\
sqrt8    & 8  & 4  & 1  & 0.09    & 0.00   \\
rd84     & 8  & 4  & 3  & 0.17    & 0.04   \\
root     & 8  & 5  & 2  & 0.16    & 0.01   \\
squar5   & 5  & 8  & 4  & 0.01    & 0.00   \\
adr4     & 8  & 5  & 1  & 0.17    & 0.02   \\
dist     & 8  & 5  & 2  & 0.25    & 0.01   \\
clip     & 9  & 5  & 2  & 1.08    & 0.06   \\
cm85a    & 11 & 3  & 2  & 0.14    & 0.30   \\
pm1      & 4  & 10 & 9  & 0.01    & 0.00   \\
sao2     & 10 & 4  & 4  & 0.14    & 0.10   \\
misex1   & 8  & 7  & 6  & 0.01    & 0.00   \\
co14     & 14 & 1  & 1  & 72.83   & 45.72  \\
dc2      & 8  & 7  & 5  & 0.05    & 0.01   \\
example2 & 10 & 6  & 4  & 1.36    & 0.15   \\
inc      & 7  & 9  & 7  & 0.01    & 0.00   \\
mlp4     & 8  & 8  & 5  & 0.19    & 0.05   \\
ryy6     & 16 & 1  & 1  & 157.65  & 101.18 \\
5xp1     & 7  & 10 & 3  & 0.10    & 0.01   \\
t481     & 16 & 1  & 1  & 1717.83 & 590.60 \\
x2       & 10 & 7  & 6  & 0.05    & 0.04   \\
sqr6     & 6  & 12 & 6  & 0.05    & 0.01   \\
dk27     & 9  & 9  & 6  & 0.01    & 0.01   \\
add6     & 12 & 7  & 1  & 46.12   & 3.89   \\
cmb      & 16 & 4  & 4  & 7.12    & 97.66  \\
ex1010   & 10 & 10 & 5  & 2.89    & 6.07   \\
C7552    & 5  & 16 & 15 & 0.01    & 0.08   \\
decod    & 5  & 16 & 15 & 0.00    & 0.09   \\
dk17     & 10 & 11 & 9  & 0.02    & 0.22   \\
pcler8   & 16 & 5  & 5  & 28.11   & 40.22  \\
tial     & 14 & 8  & 5  & 1007.04 & 40.28  \\
alu4     & 14 & 8  & 5  & 1270.29 & 36.07  \\
apla     & 10 & 12 & 12 & 0.03    & 0.50   \\
f51m     & 14 & 8  & 5  & 556.45  & 39.20  \\
cu       & 14 & 11 & 11 & 0.02    & 0.75   \\
in0      & 15 & 11 & 10 & 1.46    & 24.99  \\
0410184  & 14 & 14 & 0  & 1227.84 & 1.36   \\
apex4    & 9  & 19 & 17 & 0.95    & 38.89  \\
misex3   & 14 & 14 & 14 & 160.72  & 768.98 \\
misex3c  & 14 & 14 & 7  & 327.16  & 144.87 \\
cm163a   & 16 & 13 & 9  & 625.43  & 8.07   \\
bw       & 5  & 28 & 27 & 0.03    & 0.07   \\
pdc      & 16 & 40 & 39 & 31.09   & TO     \\
spla     & 16 & 46 & 45 & 32.72   & TO     \\
ex5p     & 8  & 63 & 60 & 0.35    & TO     \\
e64      & 65 & 65 & 64 & 0.07    & TO     \\
\hline
  \end{tabularx}
\end{table}

We have implemented Algorithm~E from Section~\ref{sec:exact-cube-based} in the
RevKit program `\emph{embed\_pla}' and evaluated it as follows.  We have taken
those functions that did not lead to a timeout when determining the minimal
number of lines using the exact cube-based approach in the previous section.
Note that using that technique the DSOP expression needs to be computed before
embedding it.

Table~\ref{tbl:ex-embedding} list some selected experimental results.  The first
three columns list the name of the function together with its number of inputs
and outputs.  The fourth and fifth columns list the number of lines of the
embedding together with the run-time required for computing the DSOP,
respectively, which of course coincide with the numbers listed in
Table~\ref{tbl:ex-line-determination}.  The last column lists the run-time which
is required for the embedding.  The run-time for DSOP computation is not
included in that time.

The run-time required for embedding the PLA is in most of the cases less
compared to the time required for computing the DSOP with the exception of some
few cases.  For the four largest functions in this requirement the embedding
algorithm leads to a time-out although the DSOP could be computed efficiently.

\subsection{BDD-based Embedding}
This section summarizes the results from three different experiments that we
implemented and performed in order to evaluated the BDD-based embedding that has
been described in Section~\ref{sec:heuristic-bdd-based}.

\begin{table}[t]
  \centering
  \caption{LGSynth'93 benchmark suite}
  \label{tab:lgsynth}
  \begin{tabularx}{\linewidth}{Xrrrrr}
    \hline
    Benchmark & $n$ & $m$ & Reading & Embedding & Run-time \\
    \hline
    duke2      &   22  &   29  &          0.00  &                0.07  &      0.07  \\
    misex3     &   14  &   14  &          0.04  &                0.14  &      0.18  \\
    misex3c    &   14  &   14  &          0.01  &                0.12  &      0.13  \\
    spla       &   16  &   46  &          0.07  &                0.38  &      0.45  \\
    e64        &   65  &   65  &          0.00  &                0.20  &      0.20  \\
    apex2      &   36  &    3  &          0.12  &                0.39  &      0.51  \\
    pdc        &   16  &   40  &          0.08  &                0.62  &      0.70  \\
    seq        &   41  &   35  &          0.15  &                0.70  &      0.85  \\
    cps        &   24  &  109  &          0.02  &                0.53  &      0.55  \\
    apex1      &   45  &   45  &          0.25  &                1.23  &      1.48  \\
    apex5      &  117  &   88  &           ---  &                 ---  &        TO  \\
    ex4p       &  128  &   28  &           ---  &                 ---  &        TO  \\
    \hline
  \end{tabularx}
\end{table}
\noindent
\subsubsection{LGSynth'93 Benchmarks}
In a first experiment, the algorithm is applied to all 37~functions of the
LGSynth'93 benchmark suite.  Since the functions are represented as PLA in this
case, we have added an option to the RevKit program `\emph{embed\_pla}' to
choose between the exact cube-based and heuristic BDD-based embedding.
Table~\ref{tab:lgsynth} lists the results for the hardest instances, i.e.~the
instances which required the largest run-time.  The first three columns of the
table list the name of the benchmark, the number of input variables~$n$, and the
number of output variables~$m$.  The remaining col\-umns list run-times in
seconds for reading the benchmark and embedding it as well as the total
run-time.  Except for two functions that could not be processed due to memory
limitations, the algorithm has no problems with handling these functions.  As a
result, efficient embeddings for them have been determined for the first time.
The largest function~\emph{cps} involves $131$ inputs and outputs.  No more than
$8$~CPU seconds are required to obtain a result.

In order to underline the importance of the variable ordering as discussed in
Section~\ref{sec:calculating-lines}, we have repeated the same experiment by
keeping the natural variable ordering
\begin{displaymath}
  \def\less{\!<\!}
  \kappa_1\less\cdots\less \kappa_m\less x_1\less\cdots\less x_n\less g_1\less\cdots\less
  g_m\less\gamma_1\less\cdots\less\gamma_n.
\end{displaymath}
In this case, 22~of the 37~functions could not have been processed due to memory
limitations. For the remaining functions, an embedding was determined. However,
this included only rather small functions.

\medskip
\noindent
\subsubsection{2-level Redundancy Functions}
Besides predefined functions that are read in from a file, additional experiments have been
carried out in which the BDDs have been created using manipulation
operations in the BDD package itself.
For this purpose, BDDs representing functions which
are applied in fault tolerant systems have been considered.
More precisely, let~$p,q\in\bbbn$, then
given variables~$x_i$ and~$y_{ij}$ for~$i=1,\dots,p$ and~$j=1,\dots,q$, the
Boolean function
\[ f=\bigwedge_{j=1}^q\bigvee_{i=1}^p x_i\land y_{ij} \]
is a~\emph{2-level redundancy function}~\cite{journals/tcs/NikolskaiaN01}.
Such functions encode cascade redundancies in critical systems and can be found
in formal methods for risk assessment~\cite{RiskAssessment}.
Further, the function~$f$ is true if and only if all columns of the matrix product
\[ x\cdot Y=(x_1x_2\ldots x_p)
\begin{pmatrix} y_{11} & y_{12} & \ldots & y_{1q} \\
                y_{21} & y_{22} & \ldots & y_{2q} \\
                \vdots & \vdots & \ddots & \vdots \\
                y_{p1} & y_{p2} & \ldots & y_{pq} \end{pmatrix} \]
are positive, i.e.~when the rows of~$Y$ selected by~$x$ cover every column of
that matrix~\cite{TAOCP4A}.

Table~\ref{tab:2level-functions} shows the results for this experiment that have
been generated using the RevKit test-case `\emph{redundancy\_functions}'.  The
columns list the values for~$p$ and~$q$, the resulting number of inputs~$n$ and
outputs~$m$ for the corresponding BDD, as well as the run-time required for the
embedding.  It can be seen that the algorithm terminates within a reasonable
amount of time for BDDs with up to 100~variables.  However, if more variables
are considered, the BDDs became too large and the algorithm ran into memory
problems.  Clearly, the efficiency of the algorithm highly depends on the size
of the BDDs.  Nevertheless, also for this set of large functions, efficient
embeddings have been obtained.

\begin{table}[t]
  \centering
  \caption{2-level redundancies functions}
  \label{tab:2level-functions}
  \begin{tabularx}{\linewidth}{XXrrr}
    \hline
    Rows $p$ & Columns $q$ & $n$ & $m$ & Run-time \\
    \hline
    5 & 5 & 30 & 1 & 0.06 \\
    6 & 6 & 42 & 1 & 0.79 \\
    7 & 7 & 56 & 1 & 6.80 \\
    8 & 8 & 72 & 1 & 77.98 \\
    9 & 9 & 90 & 1 & 1057.56 \\
    10 & 10 & 110 & 1 & 9615.86 \\
    11 & 10 & 112 & 1 & MO \\
    10 & 11 & 120 & 1 & MO \\
    11 & 11 & 132 & 1 & MO \\
    \hline
  \end{tabularx}
\end{table}

\medskip
\noindent
\subsubsection{Restricted Growth Sequences}
Similarly, another experiment has been conducted on functions representing restricted growth sequences
which should be embedded as reversible functions.
More precisely, a permutation~$\{1,\dots,p\}$ into disjoint subsets can efficiently be
represented by a string sequence $a_1,\dots,a_p$ of non-negative integers such
that~$a_1=0$ and
\[
a_{j+1}\le1+\max(a_1,\dots,a_j)\text{ for }1\le j<p.
\]
This sequence is called a~\emph{restricted growth sequence} and elements~$j$
and~$k$ belong to the same subset of the partition if and only if
$a_j=a_k$~\cite{journals/cacm/Hutchinson63,TAOCP4A}.

\begin{table}[t]
  \centering
  \caption{Restricted growth sequences}
  \label{tab:rgrowth}
  \begin{tabularx}{\linewidth}{Xrrr}
    \hline
    \multicolumn{1}{l}{Sequence length $p$} & $n$ & $m$ & \multicolumn{1}{l}{Run-time} \\
    \hline
     5 &   15 & 1 &     0.00 \\
    10 &   55 & 1 &     0.02 \\
    15 &  120 & 1 &     0.17 \\
    20 &  210 & 1 &     0.86 \\
    25 &  325 & 1 &     3.04 \\
    30 &  465 & 1 &     9.13 \\
    35 &  630 & 1 &    23.80 \\
    40 &  820 & 1 &    60.26 \\
    45 & 1035 & 1 &   139.71 \\
    50 & 1275 & 1 &   295.93 \\
    55 & 1540 & 1 &   566.00 \\
    60 & 1830 & 1 &  1029.67 \\
    65 & 2145 & 1 &  1802.98 \\
    70 & 2485 & 1 &  2966.90 \\
    75 & 2850 & 1 &  4811.23 \\
    80 & 3240 & 1 &       TO \\
    \hline
  \end{tabularx}
\end{table}
Table~\ref{tab:rgrowth} lists the results when applying the heuristic embedding
algorithm to BDDs representing these restricted growth sequences for different
sequence lengths~$p$.  The experiment has been implemented in the RevKit
test-case `\emph{restricted\_growth\_sequence}'.  The col\-umns list the length,
number of inputs and outputs, and the total run-time in seconds.  It can be seen
that here even functions with more than 400~variables can be handled within a
reasonable amount of run-time.

\section{Conclusions}
\label{sec:conclusions}
Significant progress has been made in the synthesis of reversible circuits.  In
particular, scalability has intensively been addressed. However, no solution was
available thus far that embeds large irreversible functions into reversible
ones.  In this work, we have investigated this problem extensively.  We showed
that this problem is \coNP-hard and thus intractable.  We then described
approaches both for determining the number of lines and for embedding an
irreversible function.  Sum-of-products and binary decision diagrams have been
used as function representations in these approaches and also both exact
approaches and heuristics have been presented.  For the first time, this enabled
the determination of compact embeddings of functions containing hundreds of
variables.

Future work includes the application of the proposed embedding scheme to
scalable synthesis approaches for rever\-sible functions for which thus far no
embedding has been available~\cite{STDD:14,DBLP:conf/aspdac/SoekenWHPD12}.
Further, there are some interesting open problems that resulted from the
research presented in this paper:
\begin{enumerate}
\item It would be good to have an approach for approximating the number of
  additional lines which guarantees not to give an under approximation.
\item It is interesting whether one can find an embedding for a general PLA
  representation, which may contain overlapping input cubes.
\item An exact embedding approach based on the exact BDD-based method for
  determining the minimal number of additional lines would allow for an
  embedding method that can work directly on BDDs and does not necessarily
  require a PLA representation.
\end{enumerate}
Overall, an important open problem in reversible circuit synthesis has been
solved by providing solutions to embed large irreversible functions.  Also, many
interesting new open problems are posed for future research on this topic.

\section{Acknowledgments}
The authors wish to thank Stefan Göller for many interesting discussions and his
contribution in proving the lower bounds for the embedding problem.

\end{document}